\documentclass[12pt,a4paper]{article}

\usepackage{amsmath}

\usepackage{times}
\usepackage{bm}
\usepackage{natbib}

\usepackage[plain,noend]{algorithm2e}

\usepackage{amssymb}
\usepackage{graphicx}
\usepackage{color}

\usepackage[width=1.085\textwidth,font=small]{caption}

\newcommand\argmax{\operatornamewithlimits{argmax}}

\newtheorem{theorem}{Theorem}
\newtheorem{proposition}{Proposition}
\newtheorem{corollary}{Corollary}

\begin{document}

\title{Fast \emph{DD}-classification of functional data}

\author{Karl Mosler$^{\ast}$ \and Pavlo Mozharovskyi$^{\ast}$ \and \\
\small{$^{\ast}$Statistics and Econometrics, Universit{\"a}t zu K{\"o}ln}\\
\small{Albertus Magnus Platz, 50923 K{\"o}ln, Germany}\\
\indent\\
3$^{rd}$ version
}
\date{September 28, 2015}
\maketitle

\abstract{
A fast nonparametric procedure for classifying functional data is introduced. It consists of a two-step transformation of the original data  plus a classifier operating on a low-dimensional space. The functional data are first mapped into a finite-dimensional location-slope space and then transformed by a multivariate depth function into the $DD$-plot, which is a subset of the unit square. This transformation yields a new notion of depth for functional data.
Three alternative depth functions are employed for this, as well as two rules for the final classification in $[0,1]^2$.
The resulting classifier has to be cross-validated over a small range of parameters only, which is restricted by a Vapnik-Chervonenkis bound. The entire methodology does not involve smoothing techniques, is completely nonparametric and allows to achieve Bayes optimality under standard distributional settings. It is robust, efficiently computable, and has been implemented in an R environment. Applicability of the new approach
is demonstrated by simulations as well as by a benchmark study.
}


\section{Introduction}\label{sec:introduction}

The problem of classifying objects that are represented by functional data arises in many fields of application like biology, biomechanics, medicine and economics. \cite{RamsayS05} and \cite{FerratyV06}
contain broad overviews of functional data analysis and the evolving field of classification.
At the very beginning of the 21st century many classification approaches have been extended from multivariate to functional data: linear discriminant analysis \citep{JamesH01}, kernel-based classification \citep{FerratyV03}, $k$-nearest-neighbours classifier \citep{BiauBW05}, logistic regression \citep{LengM06}, neural networks \citep{FerreV06}, support vector machines \citep{RossiV06}.
Transformation of functional data into a finite setting is done by using principal and independent \citep{HuangZ06} component analysis, principal coordinates \citep{HallPP01}, wavelets \citep{WangRM07} or functions of very simple and interpretable structure \citep{TianJ10}, or some optimal subset of initially given evaluations \citep{FerratyHV10,DelaigleHB12}.

Generally, functional data is projected onto a finite dimensional space in two ways: by either fitting some finite basis or using functional values at a set of discretization points.
The first approach accounts for the entire functional support, and the basis components can often be well interpreted.
However, the chosen basis is not \textit{per se} best for classification purposes, e.g., Principal Component Analysis (PCA) maximizes dispersion but does not minimize classification error. Moreover, higher order properties of the functions, which are regularly not incorporated, may carry information that is important for classification \citep[see][for discussion]{DelaigleH12}.
The second approach appears to be natural as the finite-dimensional space is directly constructed from the observed values. But any selection of discretization points restricts the range of the values regarded, so that some classification-relevant information may be lost. Also, the data may be given at arguments of the functions that are neither the same nor equidistant nor enough frequent. Then some interpolation is needed and interpolated data instead of the original one are analyzed. Another issue is the way the space is synthesized.
If it is a heuristic \citep[as in][]{FerratyHV10}, a well classifying configuration of discretization points may be missed. (To see this, consider three discretization points which jointly discriminate well in ${\mathbb R}^3$ constructed from their evaluations, but which cannot be chosen subsequently because each of them has a relatively small discrimination power compared to some other available discretization points.)
To cope with this problem, \cite{DelaigleHB12} consider (almost) all sets of discretization points that have a given cardinality; but this procedure involves an enormous computational burden, which restricts its practical application to rather small data sets.

\cite{LopezPR06}, \cite{CuevasFF07}, \cite{CuestaANR10} and \cite{SgueraGL13} have introduced nonparametric notions of data depth for functional data classification \citep[see also][]{CuestaAFBOF14}.
A data depth measures how close a given object is to an - implicitly given - center of a class of objects; that is, if objects are functions, how central a given function is in an empirical distribution of functions.

Specifically, the band depth \citep{LopezPR06}
of a function $\mathbf{x}$ in a class $X$ of functions indicates the relative frequency of $\mathbf{x}$ lying in a band shaped by any $J$ functions from $X$, where $J$ is fixed.
\cite{CuevasFF07} examine five functional depths for tasks of robust estimation and supervised classification: the {\em integrated depth} of \cite{FraimanM01}, which averages univariate depth values over the function's domain; the {\em $h$-mode depth}, employing a kernel; the {\em random projection depth}, taking the average of univariate depths in random directions; and the {\em double random projection depths} that include first derivatives and are based on bivariate depths in random directions. \cite{CuestaANR10} classify the Berkeley growth data \citep{TuddenhamS54}
by use of the random Tukey depth \citep{CuestaANR08}. \cite{SgueraGL13} introduce a functional spatial depth and a kernelized version of it.
\cite{CuestaAFBOF14} extensively study applications of the $DD$-plot to functional classification, employing a variety of functional depths.

There are several problems connected with the depths mentioned above.
First, besides double random projection depth, the functions are treated as multivariate data of infinite dimension. By this, the development of the functions, say in time, is not exploited. These depth notions are invariant with respect to an arbitrary rearrangement of the function values.
Second, several of these notions of functional depth break down in standard distributional settings, i.e.\ the depth functions vanish almost everywhere \citep[see][]{ChakrabortyC12,KuelbsZ13}.
Eventually, the depth takes empirical zero values if the function's hyper-graph has no intersection with the hypo-graph of any of the sample functions or vice versa, which is the case for both half-graph and band depths and their modified versions, as well as for the integrated depth.
If a function has zero depth with respect to each class it is mentioned as an \emph{outsider}, because it cannot be classified immediately and requires an additional treatment \citep[see][]{LangeMM12a,MozharovskyiML13}.

\subsection{Two benchmark problems}
Naturally, the performance of a classifier and its relative advantage over alternative classifiers depend on the actual problem to be solved.
Therefore we start with two benchmark data settings, one fairly simple and the other one rather involved.

\begin{figure*}
    \centering
    \includegraphics[keepaspectratio=true,scale=0.4]{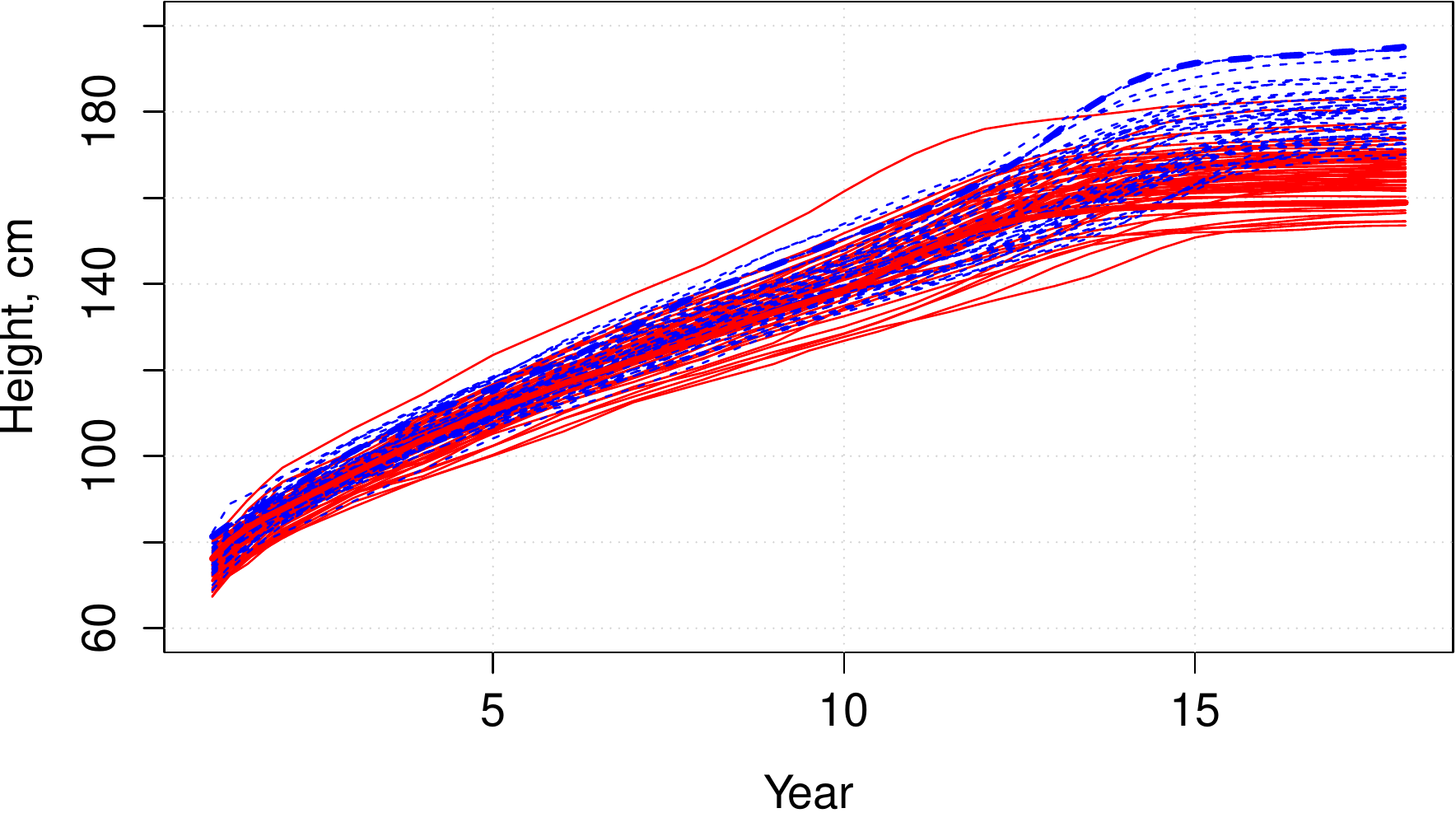}
    \includegraphics[keepaspectratio=true,scale=0.4]{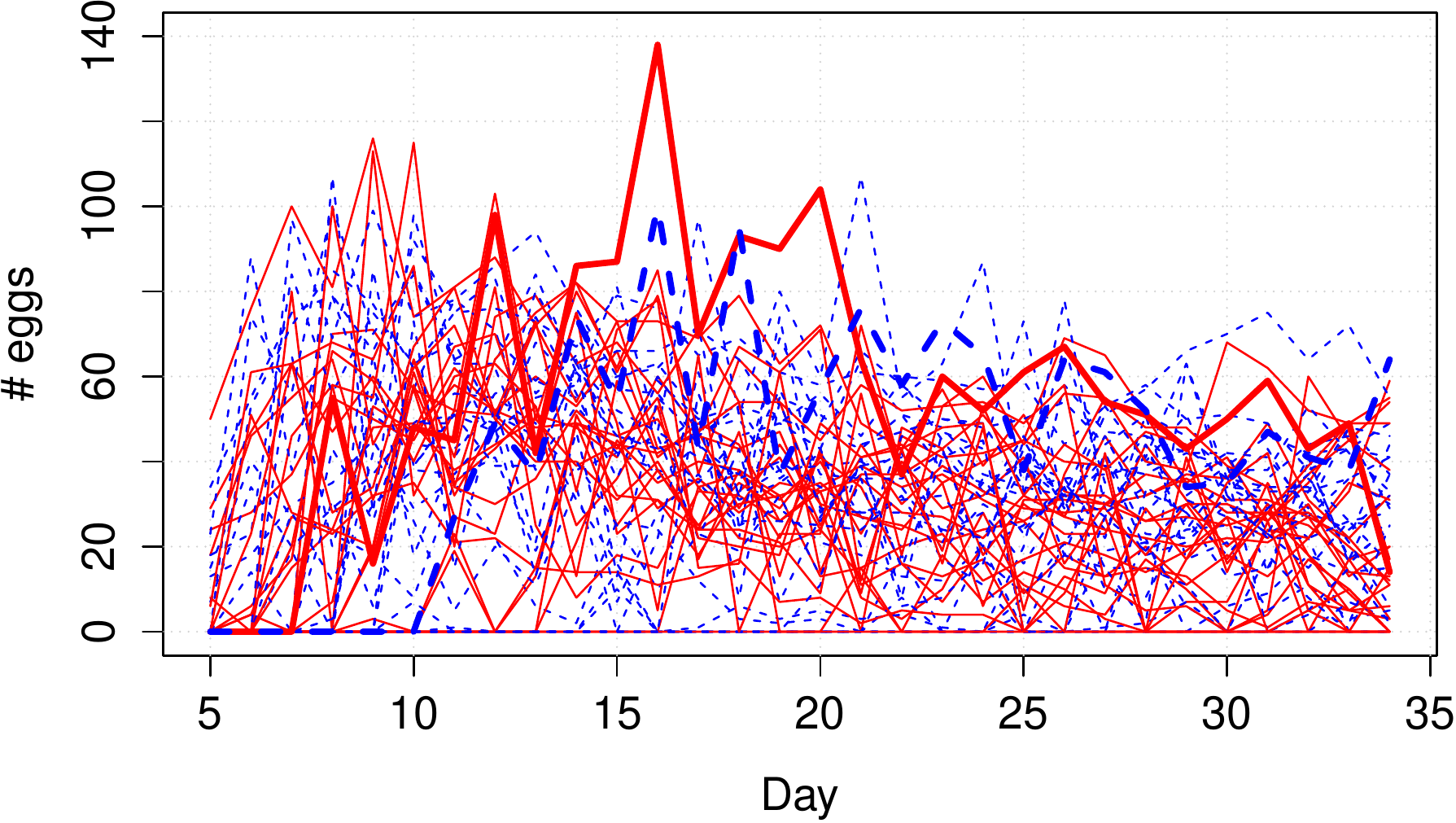}
	\caption{Growth of 54 girls (red) and 39 boys (blue) (left); 35 observations of medflies fertility (right). The fat curve indicates a single member of each class.}
    \label{fig:growthmedflies}
\end{figure*}

First, a popular benchmark set is the \emph{growth data} of the Berkeley Growth Study \citep{TuddenhamS54}.
It comprises the heights of 54 girls and 39 boys measured at 31 non-equally distant time points, see Figure~\ref{fig:growthmedflies}, left.
This data was extensively studied in \cite{RamsayS05}. After having become a well-behaved classic it has been considered as a benchmark in many works on supervised and unsupervised classification, by some authors also
in a data depth context.
Based on the generalized band depth \cite{LopezPR06} introduce a trimming and weighting of the data and classify them by
their (trimmed) weighted average distance to observation or their distance to the trimmed mean.
\cite{CuevasFF07} classify the \emph{growth data} by their maximum depth, using the five above mentioned projection-based depths and $kNN$ as a reference \citep[see also][]{CuestaANR10}.
Recently, \cite{SgueraGL13} apply functional spatial depth to this data set, and \cite{CuestaAFBOF14} perform $DD$-plot classification.

Second, we consider a much more difficult task: classifying the \emph{medflies data} of \cite{CareyLMWC98}, where 1000 30-day (starting from the fifth day) egg-laying patterns of Mediterranean fruit fly females are observed. The classification task is to explain longevity by productivity. For this a subset of 534 flies living at least 34 days is separated into two classes: 278 living 10 days or longer after the 34-th day (long-lived), and 256 those have died within these 10 days (short-lived), which are to be distinguished using daily egg-laying sequences (see Figure~\ref{fig:growthmedflies}, right, with the linearly interpolated evaluations). This task is taken from \cite{MuellerS05}, who demonstrate that the problem is quite challenging and cannot be satisfactorily solved by means of functional linear models.

\subsection{The new approach}
We shall introduce a new methodology for supervised functional classification covering the mentioned issues and validate it with the considered real data sets as well as with simulated data. Our approach involves a new notion of functional data depth. It is completely non-parametric, oriented to work with raw and
irregularly sampled functional data, and does not involve heavy computations.
It extends the idea of the location-slope depth \citep[see][]{MoslerP12}, and employs the depth-to-depth classification technique proposed by \cite{LiCAL12}, the $DD\alpha$-classification of \cite{LangeMM12a} and the depth-based $kNN$, first suggested by \cite{Vencalek11}.

Clearly, as any statistical methodology for functional data, our classification procedure has to map the relevant features of the data to some finite dimensional setting.
For this we map the functional data to a finite-dimensional location-slope space, where each function is represented by a vector consisting of integrals of its levels (`location') and first derivatives (`slope') over $L$ {respectively} $S$ equally sized subintervals.

We interpolate the functions linearly; hence their levels are integrated as piecewise-linear functions, and the derivatives as piecewise constant ones. Then we classify the data within the location-slope space using a proper depth-based technique.
We restrict $L+S$ by a Vapnik-Chervonenkis bound and determine $L$ and $S$ by cross-validation of the final classification error, which is very quickly done.
Thus, the loss of information caused by the finite-dimensional representation of data, which cannot be avoided, is measured by the final error of our classification procedure and minimized. This contrasts standard approaches using PCA projections or projections on spline bases, which penalize the loss of information with respect to a functional subspace but not to the ability of discrimination.

To construct the classification rule, the training classes in $(L,S)$-space are transformed to a depth-depth plot \citep[$DD$-plot, see][]{LiCAL12}, which is a two-dimensional set of points, each indicating the depth of a data point regarding the two training classes.
 Finally, the classes are separated on the $DD$-plot by either applying a $k$-nearest-neighbor ($kNN$) rule \citep{Vencalek11} or the $\alpha$-procedure
\citep{LangeMM12a}.

The resulting functional data depth, called \emph{integral location-slope depth} has a number of advantages. It is based on a linear mapping which does not introduce any spurious information and {can preserve} Bayes optimality under standard distributional assumptions.
Already in the first step, the construction of the finite-dimensional space aims at the minimization of an overall goal function, \emph{viz.} the misclassification rate.
The simplicity of the mapping allows for its efficient computation.

The procedure has been implemented in the R-package {\tt ddalpha}.


\subsection{Overview}
The rest of the paper is organized as follows: Section~\ref{sec:depthclass} reviews the use of data depth techniques in classifying finite-dimensional objects. Section~\ref{sec:depthtransform} presents the new two-step representation of functional data, first in a finite-dimensional Euclidean space (the location-slope space), and then in a depth-to-depth plot ($DD$-plot). In Section~\ref{sec:ddclassification} we briefly introduce two alternative classifiers that operate on the $DD$-plot, a nearest-neighbor procedure and the $\alpha$-procedure. Section~\ref{sec:props} treats some theoretical properties of the developed classifiers. Section~\ref{sec:lss} provides an approach to bound and select the dimension of the location-slope space. In Section~\ref{sec:experiments} our procedure is applied to simulated data and compared with other classifiers.
Also, the above two real data problems are solved and the computational efficiency
of our approach is discussed. Section~\ref{sec:conclusions} concludes.
Implementation details and additional experimental results are collected in the Appendix.

\section{Depth based approaches to classification in $\mathbb{R}^d$}\label{sec:depths}
\label{sec:depthclass}
For data in Euclidean space $\mathbb{R}^d$ many special depth notions have been proposed in the literature; see, e.g., \cite{ZuoS00} for definition and properties.
Here we mention three depths, Mahalanobis, spatial and projection depth.
These depths are everywhere positive. Hence they do not produce outsiders, that is, points having zero depth in both training classes.
This no-outsider property appears to be essential in obtaining functional depths for classification.

For a point $\mathbf{y}\in \mathbb{R}^d$ and a random vector $Y$ having an empirical distribution on $\{\mathbf{y}_{1}, \dots ,\mathbf{y}_{n}\}$ in $\mathbb{R}^{d}$ the \emph{Mahalanobis depth} \citep{Mahalanobis36} of $\mathbf{y}$ w.r.t. $Y$ is defined as
\begin{equation}\label{eqn:MahDepth}
D^{Mah}(\mathbf{y}|Y) = \bigl(1 + (\mathbf{y} - \mu_{Y})^{\prime}\Sigma_{Y}^{-1}(\mathbf{y} - \mu_{Y})\bigr)^{-1},
\end{equation}
where $\mu_{Y}$ measures the location of $Y$, and $\Sigma_{Y}$ the scatter.

The \emph{affine invariant spatial depth} \citep{VardiZ00,Serfling02} of $\mathbf{y}$ regarding $Y$ is defined as
\begin{equation}\label{eqn:sptDepth}
D^{Spt}(\mathbf{y}|Y) = 1 - \| E_Y \left[ v\bigl(\Sigma_{Y}^{-1/2}(\mathbf{y} - Y)\bigr) \right]\|\,,
\end{equation}
where $\|\cdot\|$ denotes the Euclidean norm, $v(\mathbf{w})=\|\mathbf{w}\|^{-1}\mathbf{w}$ for $\mathbf{w}\ne \mathbf{0}$ and $v(\mathbf{0})=\mathbf{0}$, and $\Sigma_{Y}$ is the covariance matrix of $Y$. $D^{Mah}$ and $D^{Spt}$ can be efficiently computed in $\mathbb{R}^{d}$.

The \emph{projection depth} \citep{ZuoS00} of $\mathbf{y}$ regarding $Y$ is given by
\begin{equation}\label{eqn:prjDepth1}
D^{Prj}(\mathbf{y}|Y) = \inf_{\mathbf{u} \in S^{d-1}}\bigl(1 + O^{Prj}(\mathbf{y}|Y, \mathbf{u})\bigr)^{-1},
\end{equation}
with
\begin{equation}\label{eqn:prjDepth2}
O^{Prj}(\mathbf{y}|Y, \mathbf{u}) = \frac{\left|\mathbf{y}^\prime\mathbf{u} - m(Y^\prime\mathbf{u})\right|}{MAD(Y^\prime\mathbf{u})}\,,
\end{equation}
where $m$ denotes the univariate median and $MAD$ the median absolute deviation from the median.
Exact computation of $D^{Prj}$ is, in principle, possible \citep{LiuZ12} but practically infeasible when $d>2$.
Obviously, $D^{Prj}$ is approximated from above by calculating the minimum of univariate projection depths in random directions.
However, as $D^{Prj}$ is piece-wise linear (and, hence, attains its maximum on the edges of the direction cones of constant linearity), a randomly chosen direction yields the exact depth value with probability zero.
For a sufficient approximation one needs a huge number of directions, each of which involves the calculation of the median and MAD of a univariate distribution.

To classify objects in Euclidean space $\mathbb{R}^d$, the existing literature employs depth functions in principally two ways:
\begin{enumerate}
\item Classify the original data by their maximum depth in the training classes.
\item Transform the data by their depths into a low-dimensional depth space, and classify them within this space.
\end{enumerate}

\emph{Ad} 1:
\cite{GhoshC05b} propose the \emph{maximum depth classifier}, which assigns an object to the class in which it has maximum depth.
In its naive form this classifier yields a linear separation.
Maximum-depth classifiers have a plug-in structure;
their scale parameters need to be tuned (usually by some kind of cross-validation) over the whole learning process. For this, \cite{GhoshC05b} combine the naive maximum-depth classifier with an estimate of the density. A similar approach is pursued with projection depth in \cite{DuttaG12a} and $l_p$ depth in \cite{DuttaG12b}, yielding competitive classifiers.

\emph{Ad} 2: Depth notions are also used to reduce the dimension of the data. \cite{LiCAL12} employ the $DD$-plot, which represents all objects by their depth in the two training classes, that is, by points in the
unit square. (The same for $q$ training classes in the $q$-dimensional unit cube.) To solve the classification task, some separation rule has to be constructed in the unit square. \cite{LiCAL12} minimize the \emph{empirical risk}, that is the average classification error on the training classes, by smoothing it with a logistic sigmoid function and, by this, obtain a polynomial separating rule. They show that their approach (with Mahalanobis, projection and other depths) asymptotically achieves the optimal Bayes risk if the training classes are strictly
unimodal elliptically distributed. However, in practice the choice of the smoothing constant and non-convex optimization, potentially with many local minima, encumber its application. In \cite{LangeMM12a} these problems are addressed via the $\alpha$-procedure, which is very fast and speeds up the learning phase enormously. Use of other procedures in the $DD$-plot is discussed in
\cite{CuestaAFBOF14}.


\section{A new depth transform for functional data}\label{sec:depthtransform}

Let ${\mathcal F}$ be the space of real functions, defined on a compact interval,
which are continuous and smooth at all points besides a finite set, and let $\mathcal{F}$ be endowed with the supremum norm.
The data may be given either as observations of complete functions in ${\cal F}$ or as functional values at some discretization points, in general neither equidistant nor common ones.
If the functional data is given in discretized form, it is usually interpolated by splines of some order
\citep[see][]{RamsayS05}, so that sufficiently smooth functions are obtained. Here we use linear interpolation (that is splines of order 1) for the following reasons. Firstly, with linear interpolation, neither the functions nor their derivatives need to be smoothed. Thus almost any raw data can be handled easily. E.g., the \emph{ medflies data}, as to the egg-laying process, is naturally discrete; see Figure~\ref{fig:growthmedflies} (right). Secondly, higher order splines increase the computational load, especially when the number of knots or the smoothing parameter are to be determined as part of the task.
Thirdly, splines of higher order may introduce spurious information.

We construct a depth transform as follows. In a first step, the relevant features of the functional data are extracted
from ${\cal F}$ into a finite-dimensional Euclidean space $\mathbb{R}^{L+S}$, which we call \emph{the location-slope space} (Section~\ref{ssec:lstransform}). Then an $(L+S)$-dimensional depth is applied to the transformed data
yielding a $DD$\emph{-plot} in the unit square (Section~\ref{ssec:ddtransform}), which represents the two training classes. Finally the separation of the training classes as well as the classification of new data is done on the $DD$-plot (Section~\ref{sec:ddclassification}).

\subsection{The location-slope transform}\label{ssec:lstransform}
\begin{figure*}
    \centering
    \includegraphics[keepaspectratio=true,scale=0.375]{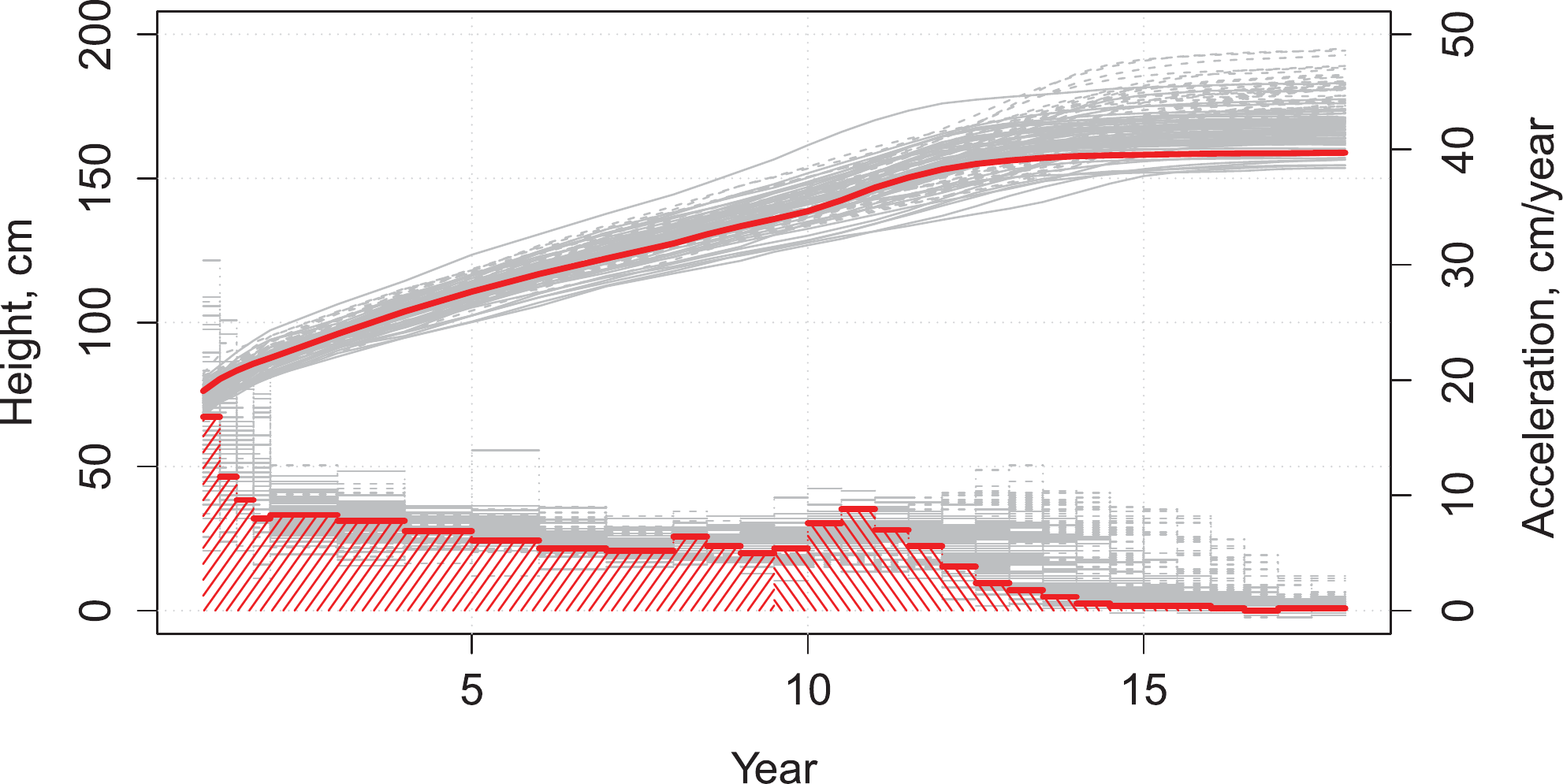}
    \includegraphics[keepaspectratio=true,scale=0.375]{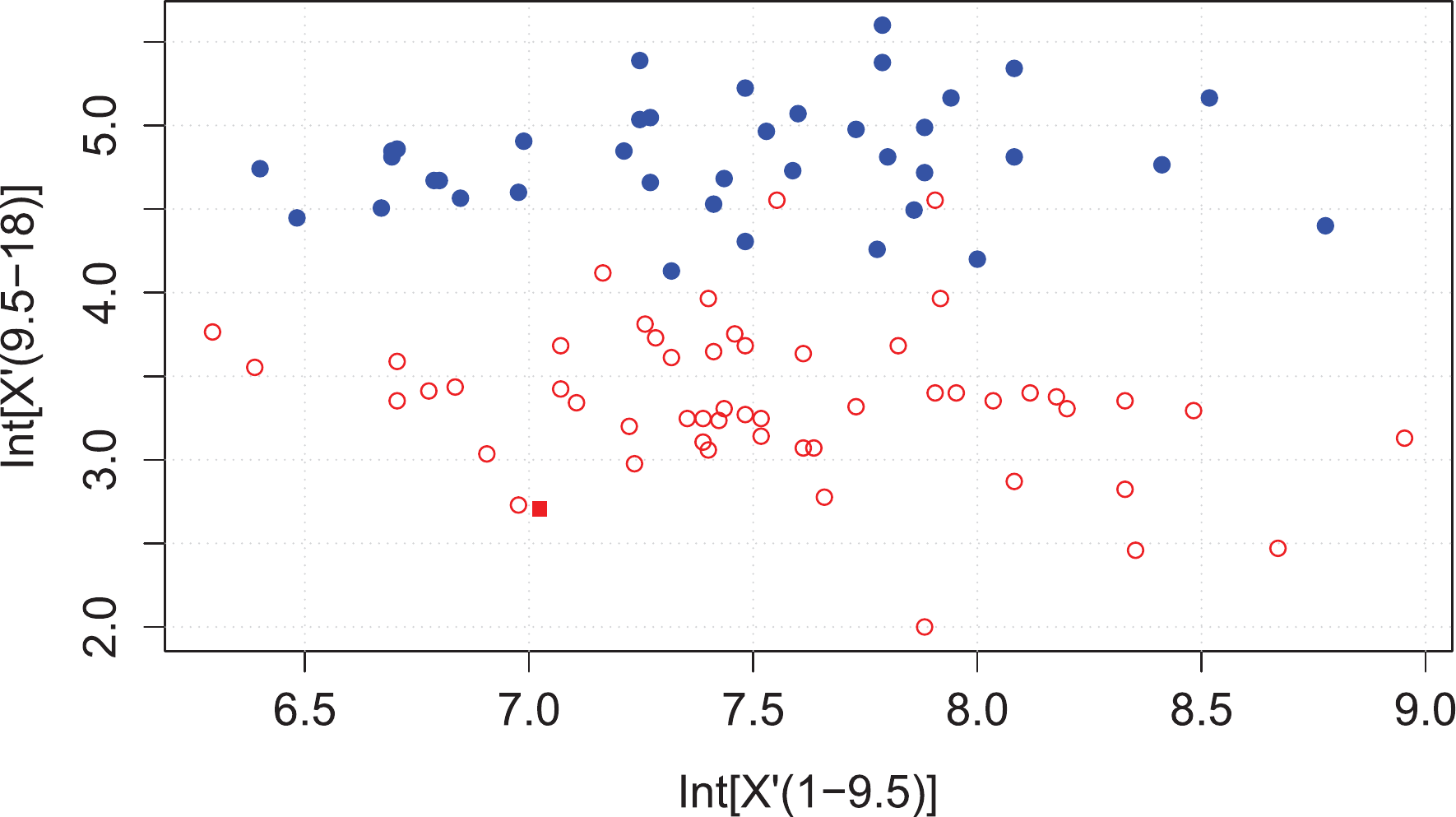}
    \caption{Example: Transformation of a single growth function into the integrated location-slope space with $L=0$ and $S=2$. Function(s) and first derivative(s) (left); corresponding two-dimensional space $(0,2)$ (right).}
    \label{fig:growth2}
\end{figure*}

We consider two classes of functions in ${\cal F}$, $X_0=\{\mathbf{\tilde x}_{1}, \dots ,\mathbf{\tilde x}_{m}\}$ and $X_1=\{\mathbf{\tilde x}_{m+1}, \dots,\mathbf{\tilde x}_{m+n}\}$, which are given as measurements at ordered points $t_{i1}\le t_{i2}\le ... \le t_{ik_i}$, $i=1,...,m+n$,
\begin{equation}
\left[\mathbf{\tilde x}_{i}(t_{i1}), \mathbf{\tilde x}_{i}(t_{i2}), ..., \mathbf{\tilde x}_{i}(t_{ik_i})\right]\,.
\end{equation}
Assume w.l.o.g. $\min_i{t_{i1}}=0$ and let $T=\max_i{t_{ik_i}}$.
From $\mathbf{\tilde x}_{i}$ a function $\mathbf{x}_{i}:[0,T]\to \mathbb{R}$ is obtained as follows.
Connect the points $(t_{ij},\, \mathbf{\tilde x}_{i}(t_{ij})), j=1,\dots, k_i,$ with line segments and
set $\mathbf{x}_{i}(t)=\mathbf{\tilde x}_{i}(t_{i1})$ when $0\le t\le t_{i1}$ and $\mathbf{x}_{i}(t)=\mathbf{\tilde x}_{i}(t_{ik_i})$ when $t_{ik_i}\le t\le T$.
By this, the data become piecewise linear functions on $[0,T]$, and their first derivatives become piecewise constant ones, see Figure~\ref{fig:growth2}, left.

A finite-dimensional representation of the data is then constructed by integrating the interpolated functions over $L$ subintervals (location) and their derivatives over $S$ subintervals (slope), see Figure~\ref{fig:growth2} (left). Thus, our location-slope space has dimension $L+S$.
It delivers the following transform,
\begin{eqnarray} \label{average}
& \mathbf{x}_{i} \quad \longmapsto \quad \mathbf{y}_{i} = & \Bigl[\int_0^{T/L} \mathbf{x}_{i}(t) dt, \dots,  \int_{T(L-1)/L}^{T} \mathbf{x}_{i}(t) dt, \\
&& \int_0^{T/S} \mathbf{x'}_{i}(t) dt, \dots,  \int_{T(S-1)/S}^{T} \mathbf{x'}_{i}(t) dt \Bigr]\,. \nonumber
\end{eqnarray}
That is, the $L+S$ average values and slopes constitute a point $\mathbf{y}_{i}$ in $\mathbb{R}^{L+S}$.
Either $L$ or $S$ must be positive, $L+S\ge 1$. In case $L=0$ or $S=0$ the formula (\ref{average}) is properly modified.
Put together we obtain a composite transform $\phi:\mathcal{F} \to \mathbb{R}^{L+S}$,
\begin{equation}\label{LStransform}
\mathbf{\tilde x}_{i} \quad \mapsto \quad \left[\mathbf{\tilde x}_{i}(t_{i1}), \dots, \mathbf{\tilde x}_{i}(t_{ik_i})\right] \quad \mapsto \quad
\mathbf{x}_{i} \quad \mapsto \quad \mathbf{y}_i\,,
\end{equation}
which we call the \textit{location-slope (LS-) transform}. Note that also \cite{CuevasFF07} and \cite{CuestaAFBOF14} include derivatives in their functional depths, but in a different way.

For example, choose $L=0$ and $S=2$ for the \emph{growth data}. Then they are mapped into the location-slope space $\mathbb{R}^{L+S}=\mathbb{R}^2$, which is shown in Figure~\ref{fig:growth2} (right). Here the functions' first derivatives are averaged on two half-intervals. That is, for each function two integrals of the slope are obtained: the integral over $[1,9.5]$ and that over $[9.5,18]$. Here, the location is not incorporated at all. Figure~\ref{fig:growth2}, left, exhibits the value (height) and first derivative (velocity) of a single interpolated function, which is then represented by the average slopes on the two half-intervals, yielding the rectangular point in Figure~\ref{fig:growth2} (right).

Further, given the data, $L$ and $S$ can be chosen large enough to reflect all relevant information about the functions.
(Note that the computational load of the whole remaining procedure is linear in dimension $L+S$.)
If $L$ and $S$ are properly chosen, under certain standard distributional assumptions, the $LS$-transform preserves asymptotic Bayes optimality; see Section~\ref{sec:props} below.

Naturally, only part of these $L+S$ intervals carries the information needed for separating the two classes. \cite{DelaigleHB12} propose to determine a subset of points in $[0,T]$ based on which the training classes are optimally separated. However they do not provide a practical procedure to select these points; in applications they use cross-validation.
{Generally, we have} no prior reason to weight the intervals differently.
Therefore we use intervals of equal length, but possibly different ones for location and slope.

The question remains how many equally sized subintervals, $L$ for location and $S$ for slope, should be taken.
We will see later in Section~\ref{sec:experiments} that our classifier performs similar with the three depths when the dimension is low. In higher dimensions the projection depth cannot be computed precisely enough, so that the classifier becomes worse.
The performance is most influenced by the construction of the location-slope transform, that is, by the choice of the numbers $L$ and $S$. We postpone this question to Section~\ref{sec:lss}.

\subsection{The $DD$-transform}\label{ssec:ddtransform}
\begin{figure*}
    \centering
	\includegraphics[keepaspectratio=true,scale=0.415]{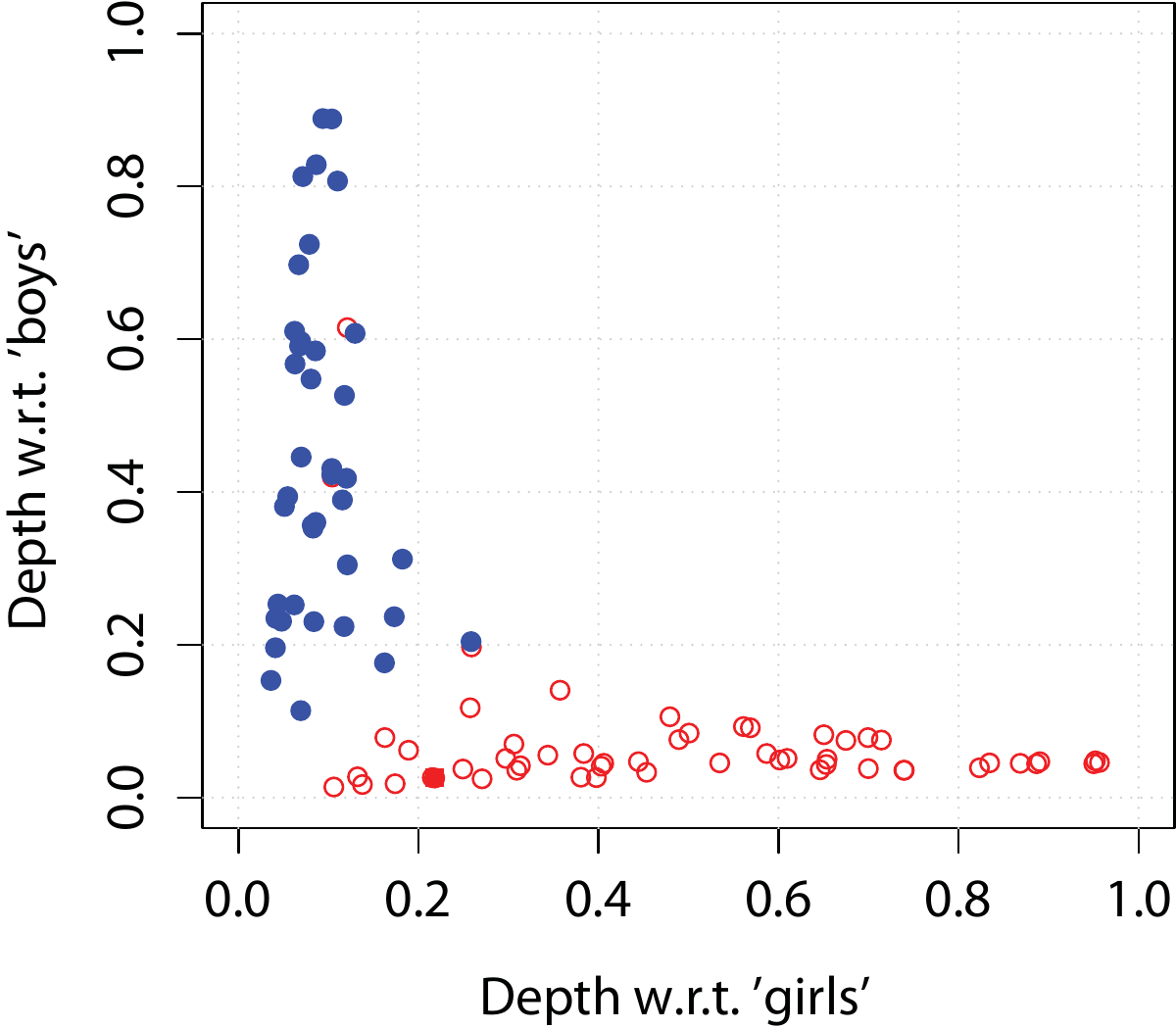}
    \includegraphics[keepaspectratio=true,scale=0.415]{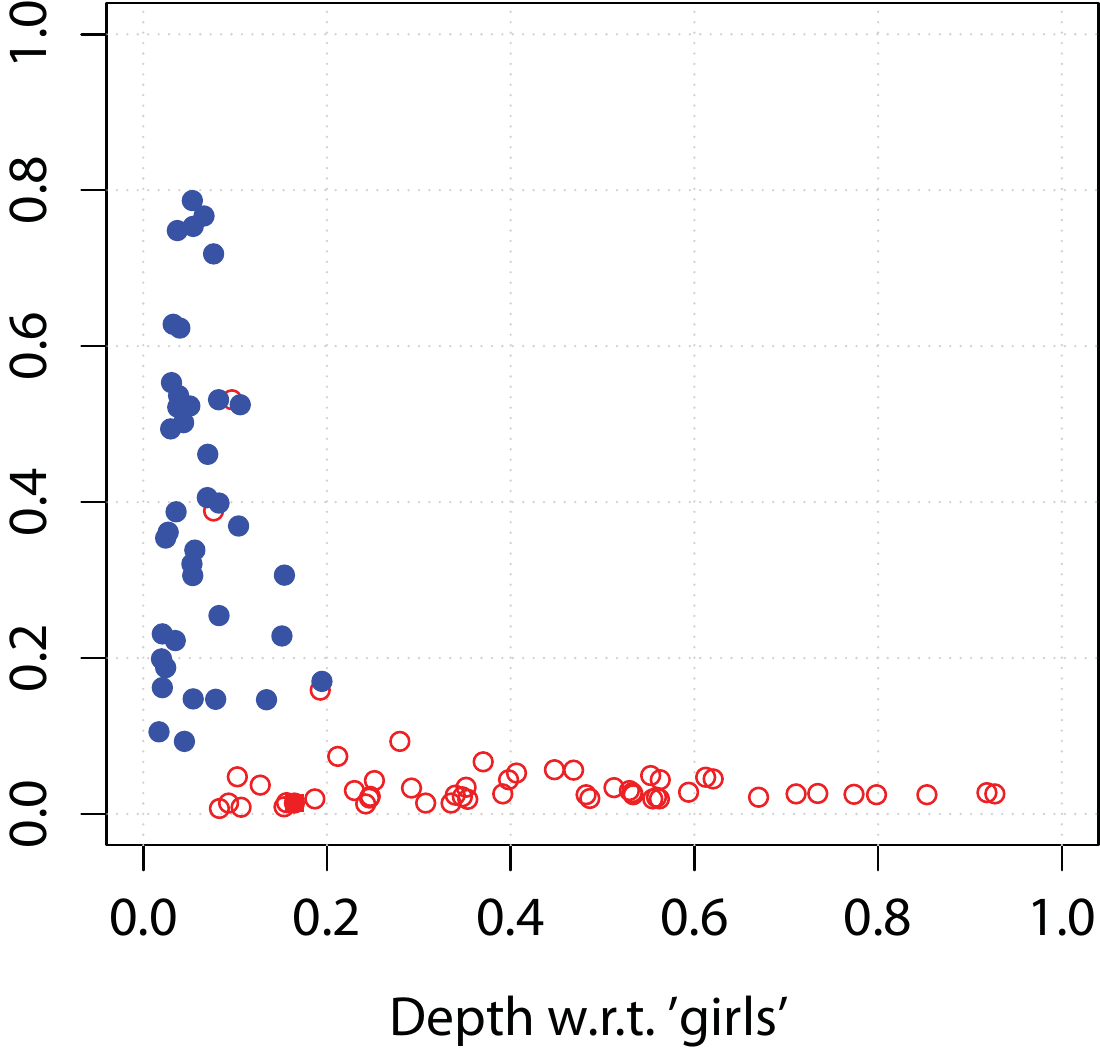}
    \includegraphics[keepaspectratio=true,scale=0.415]{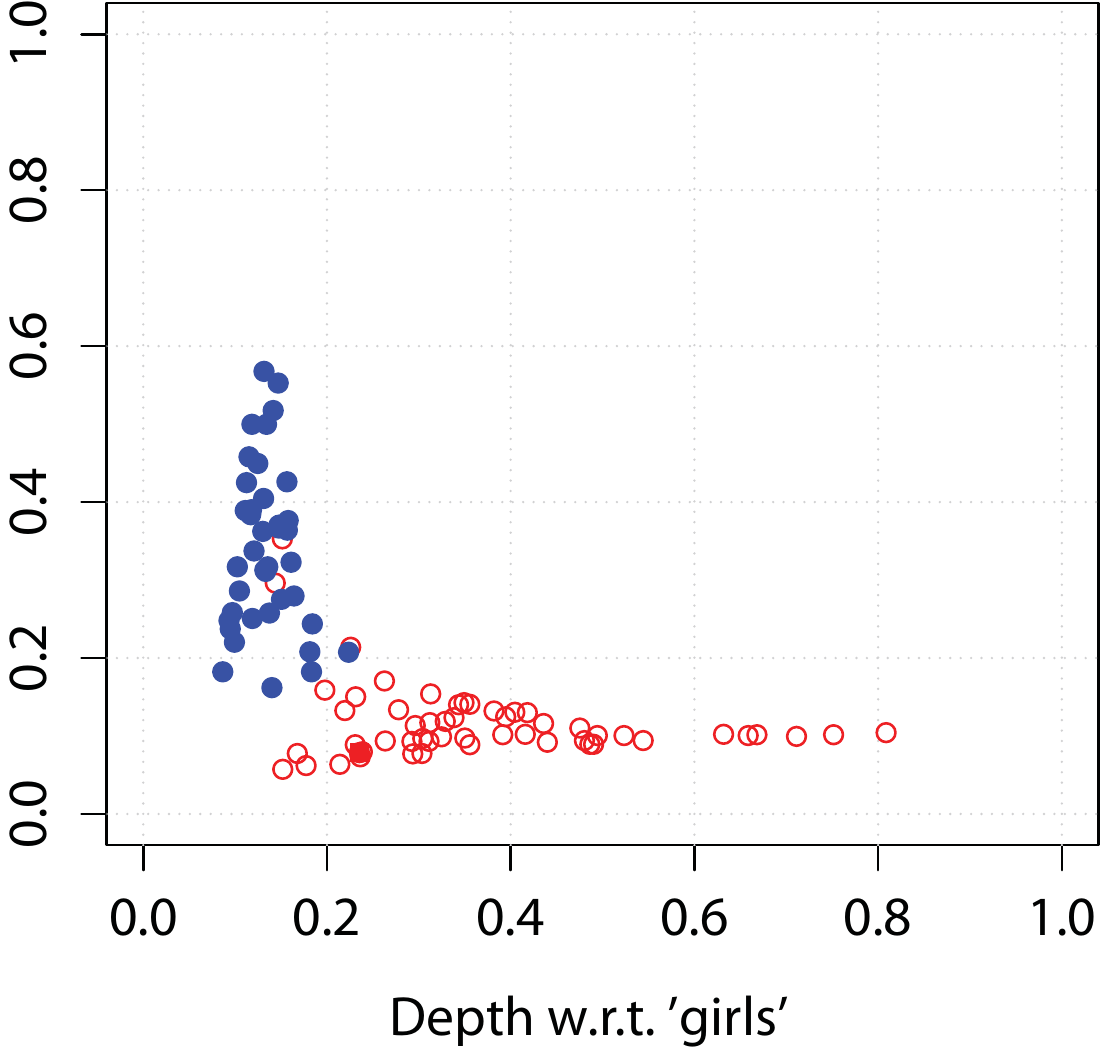}
	\caption{$DD$-plots for \emph{growth data} with $(L,S)=(0,2)$ using Mahalanobis (left), spatial (middle) and projection (right) depths.}
    \label{fig:growth3}
\end{figure*}

Denote the location-slope-transformed training classes in $\mathbb{R}^{L+S}$ by $Y_0=\{\mathbf{y}_{1}, \dots ,\mathbf{y}_{m}\}$ and $Y_1=\{\mathbf{y}_{m+1}, \dots,\mathbf{y}_{m+n}\}$. The $DD$-plot is then
\begin{eqnarray}
Z & = & \{\mathbf{z}_i=(z_{i0},z_{i1}) \, | \, z_{i0}=D^{L+S}(\mathbf{y}_{i}|Y_0),\\
&& \,\,z_{i1}=D^{L+S}(\mathbf{y}_{i}|Y_1),i=1,...,m+n\}\,.\nonumber
\end{eqnarray}
Here $D^{L+S}$ is an $(L+S)$-dimensional depth. In particular, $D^{L+S}$ may be $D^{Mah}$, $D^{Spt}$ or $D^{Prj}$, each of which does not produce outsiders.
The $DD$-plots of these three for \emph{growth data}, taking $L=0$ and $S=2$, are pictured in Figure~\ref{fig:growth3}. Clearly, in this case, almost faultless separation is achievable by drawing a straight line through the origin.

\section{$DD$-plot classification}\label{sec:ddclassification}

The training phase of our procedure consists in the following. After the training classes have been mapped from ${\cal F}$ to the $DD$-plot as described in Section~\ref{sec:depthtransform}, a selector is determined in the $DD$-plot that separates the $DD$-transformed data.
For the latter, we consider two classifiers operating on the $DD$-plot, and compare their performance.
Firstly we propose the $kNN$ classifier. It
converges to the Bayes rule under standard elliptical models;
see Section~\ref{ssec:ddplotknn}. Note that $kNN$ needs to be cross-validated on the $DD$-plot, which is computationally expensive. Therefore, secondly, we employ the $DD\alpha$-procedure, which is a very fast heuristic; see Section~\ref{ssec:ddalpha}.
Although its theoretical convergence is not guaranteed \citep[see][]{LangeMM12a}, the $DD\alpha$-classifier performs very well in applications.

\subsection{$kNN$-classification on the $DD$-plot}\label{ssec:ddplotknn}
When applied to multivariate data, (under mild assumptions) $kNN$ is known to be a consistent classifier. By multiplying all distances with the inverse covariance matrix of the pooled data an affine invariant version is obtained.
In our procedure we employ an affine-invariant $kNN$ classifier on the $DD$-plot.
It will be shown (Section~\ref{sec:props}) that, if the underlying distribution of each class is strictly unimodal elliptical with the same radial density, the $kNN$ classifier, operating on the $DD$-plot, achieves asymptotically the optimal Bayes risk.

Given a function $\mathbf{x}_0\in {\cal F}$ to be classified, we represent $\mathbf{x}_0$ (according to Section~\ref{sec:depthtransform}) as $\mathbf{y}_0\in \mathbb{R}^{L+S}$ and then as
$\mathbf{z}_0= \left(D^{L+S}(\mathbf{y}_0|Y_0),D^{L+S}(\mathbf{y}_0|Y_1)\right)$.
According to $kNN$ on the $DD$-plot $\mathbf{x}_0$ is classified as follows:
\begin{equation}\label{eqn:knn}
    class(\mathbf{x}_0)=I{\left(
    \sum^m_{i=1}I(\mathbf{z}_i\in R^{\beta(k)}_{\mathbf{z}_0}) <
    \sum^{m+n}_{i=m+1}I(\mathbf{z}_i\in R^{\beta(k)}_{\mathbf{z}_0})
    \right)}\,,
\end{equation}
where $I(S)$ denotes the indicator function of a set $S$, and $R^{\beta(k)}_{\mathbf{z}_0}$ is the neighborhood of $\mathbf{z}_0$ defined by the $k$-closest observations. We use the $L_\infty$-distance on the $DD$-plot; thus $R^{\beta(k)}_{\mathbf{z}_0}$ is the smallest rectangle centered at $\mathbf{z}_0$ that contains $k$ training observations.
In applications $k$ has to be chosen, usually by means of cross-validation.

\subsection{$DD\alpha$-classification}\label{ssec:ddalpha}
The second classification approach is the $DD\alpha$-classifier, introduced in \cite{LangeMM12a}.
It uses a projective-invariant method called the $\alpha$-procedure \citep{VasilevL98}, which is a heuristic classification procedure that iteratively decreases the empirical risk.
The $DD\alpha$-classifier employs a modified version of the $\alpha$-procedure to construct a nonlinear separator in $[0,1]^2$ that classifies the depth-represented data points.
The construction is based on depth values and the products of depth values (= \emph{D-features}) up to some degree $p$ that can be either chosen \textit{a priori} or determined by cross-validation.
In a stepwise way, linear discrimination is performed in subspaces of the feature space.
In each step a pair of $D$-features is replaced by a new $D$-feature as long as the average misclassification rate decreases and basic $D$-features are left to be replaced. For details, we refer to \cite{LangeMM12a}.

We employ three depths (Mahalanobis, spatial, and projection depths), which are positive on the whole $\mathbb{R}^{L+S}$ and thus do not produce outsiders. It is known that the $DD\alpha$-classifier is asymptotical Bayes-optimal for the location-shift model \citep[see][]{LangeMM12a} and
performs well for broad classes of simulated distributions and a wide variety of real data \citep[see][]{LangeMM12b,MozharovskyiML13}.
The main advantage of the $DD\alpha$-classifier is its high training speed, as it contains the $\alpha$-procedure which, on the $DD$-plot, has the quick-sort complexity $O\bigl((m+n)\log(m+n)\bigr)$ and proves to be very fast. The separating polynomial is constructed by space extensions of the $DD$-plot (which is of low dimension) and cross-validation.

\section{Some theoretical properties of the classifiers}\label{sec:props}

In this section, we treat optimality properties of classification procedures that operate on a {\it DD}-plot of $LS$-transformed functional data.
Besides $kNN$-classification and $\alpha$-procedure they include linear and quadratic discriminant analysis as well as maximum depth classification, all applied to the {\it DD}-plotted data. The latter procedures will be included in the comparative study of Section~\ref{sec:experiments}.
First, in Subsection~\ref{ssec:maropt}, it is shown that well-known properties regarding the Bayes optimality of classifiers for multivariate data
carry over to $LS$-transformed functional data if the discretization points are fixed and $L$ is chosen large enough.
Second, in Subsection~\ref{ssec:funopt}, we introduce a sampling scheme that evaluates the functions at an increasing number of points, and establish Bayes optimality in the Gaussian case.

Let ${\mathcal F}$ be the space of real functions, defined on the finite interval $[0,T]$, which are continuous and piecewise smooth.
Consider two stochastic processes, $\mathfrak{G}_0$ and $\mathfrak{G}_1$ on a common probability space $(\Omega, {\cal A}, P)$, whose paths are in $\mathcal{F}$ with probability 1. (For conditions on Gaussian processes that yield smooth paths, see e.g. \cite{Cambanis73}.)

\subsection{Fixed-discretization optimality}\label{ssec:maropt}

Let ${\mathcal T}=\{t_{j}|j=1,...,\ell\}\in[0,T]$ be a given finite set of discretization points,
and for each $\mathbf{\tilde v}\in{\mathcal F}$ let ${\mathbf{v}}$ be its linear interpolation based on $\mathbf{\tilde v}(t_1), \dots, \mathbf{\tilde v}(t_\ell)$, as described in Section \ref{sec:depthtransform}. Denote the space of these linear interpolations by ${\mathcal F}({\mathcal T})$.
Then ${\mathcal F}({\mathcal T})$ is a finite-dimensional linear space, which is isomorphic to $\mathbb{R}^\ell$.
 Consider two samples of independent random vectors in $\mathbb{R}^\ell $, $[\mathbf{\tilde v}_i(t_1), \dots, \mathbf{\tilde v}_i(t_\ell)]$, distributed as $F_0$, $i=1,\dots, m$, and $[\mathbf{\tilde w}_j(t_1), \dots, \mathbf{\tilde w}_j(t_\ell)]$, distributed as $F_1$, $j=1,\dots, n$, where $F_0$ and $F_1$ are the ${\mathcal T}$-marginal distributions of $\mathfrak{G}_0$ resp.\ $\mathfrak{G}_1$ in $\mathbb{R}^\ell $, and let the two samples be independent from each other.

\begin{proposition}[Bayes optimality on ${\mathcal F}({\mathcal T})$]\label{prop1}
Consider a class ${\mathcal C}$ of decision rules ${\mathbb R}^\ell\to\{0,1\}$ and assume that ${\mathcal C}$ contains a sequence whose risk converges in probability to the Bayes risk.
Then there exists a pair $(L^*,S^*)$ so that the same class of decision rules operating on $LS$-transformed data in ${\mathbb R}^{L+S}$
contains a sequence whose risk also converges in probability to the Bayes risk.
\end{proposition}

We say that a sequence of decision rules is \textit{Bayes optimal} if their risk converges in probability to the Bayes risk.
The proof of Proposition \ref{prop1} is obvious: E.g., choose $S^*=0$ and $L^*>T/\min_{i=1,...,\ell-1}\{t_{i+1} - t_i\}$. Then the Bayes optimality trivially carries over from $\mathbb{R}^{L+S}$ to $\mathbb{R}^{\ell}$, hence to ${\mathcal F}({\mathcal T})$.
 \hfill $\Box$

The Proposition allows for procedures that achieve error rates close to minimum. However note that it does not refer to Bayes optimality of classifying the underlying process, but just of classifying the  $\ell$-dimensional marginal distributions corresponding to $\mathcal{T}$.
In particular, if two processes have equal marginal distributions on ${\mathcal T}$, no discrimination is possible.

\begin{corollary}[Optimality of $QDA$ and $LDA$]\label{cor1}
Let $\mathfrak{G}_0$ and $\mathfrak{G}_1$ be Gaussian processes and the priors of class membership be equal, and let proper $L^*$ and $S^*$ be selected as above. The following rules operating on $LS$-transformed data are Bayes optimal, relative to the
${\cal T}$-marginals:
\begin{description}
  \item[(i)] quadratic discriminant analysis ($QDA$),
  \item[(ii)] linear discriminant analysis ($LDA$), if $\mathfrak{G}_0$ and $\mathfrak{G}_1$ have the same covariance function.
\end{description}
\end{corollary}
For a definition of LDA and QDA, see, e.g., \cite{DevroyeGL96}, Ch. 4.4).
Proof: As the processes are Gaussian, we obtain that
\begin{align}\label{Sigmas}
(\mathbf{v}_i(t_1), \dots, \mathbf{v}_i(t_\ell)) \sim (i.i.d.) N(\mathbf{\mu}_0, \mathbf{\Sigma}_0)\,,\quad i=1,\dots, m\,,\\
(\mathbf{w}_j(t_1), \dots, \mathbf{w}_j(t_\ell)) \sim (i.i.d.) N(\mathbf{\mu}_1, \mathbf{\Sigma}_1)\,,\quad j=1,\dots, n\,.
\end{align}
As by selecting $(L^*,S^*)$ the Bayes optimality is trivially preserved, the standard results of Fisher (see, again, \cite{DevroyeGL96}) apply; hence Corollary \ref{cor1} holds. \hfill $\Box$

The following result is taken from \cite{LangeMM12a}. Let $H$ be a hyperplane in $\mathbb{R}^\ell$, and let $pr_H$ denote the projection on $H$.  A probability distribution $F_1$ is mentioned as the \textit{mirror image regarding $H$} of another probability distribution $F_0$ in $\mathbb{R}^\ell$ if for $X\sim F_0$ and $Y\sim F_1$ it holds: $Y-pr_H(Y)$ is distributed as $-(X-pr_H(X))$.

\begin{proposition}[\cite{LangeMM12a}]\label{Prop2}
Let $F_0$ and $F_1$ be probability distributions in $\mathbb{R}^\ell$ having densities $f_0$ and $f_1$, and let $H$ be a hyperplane such that $F_1$ is the mirror image of $F_0$ with respect to $H$ and $f_0\ge f_1$ in one of the half-spaces generated by $H$. Then, based on a 50:50 independent sample from $F_0$ and $F_1$, the $DD\alpha$-procedure will asymptotically yield the linear separator that corresponds to the bisecting line of the $DD$-plot.
\end{proposition}
Notice that due to the mirror symmetry of the distributions in $\mathbb{R}^l$ the $DD$-plot is symmetrically distributed as well. Symmetry axis is the bisector, which is obviously  the result of the $\alpha$-procedure when the sample is large enough.
Then the $DD\alpha$-rule corresponds to the Bayes rule. In particular, the requirements of the proposition are satisfied if $F_0$ and $F_1$ are mirror symmetric and unimodal.

A stochastic process $\{X_t\}$ is mentioned as a \textit{strictly unimodal elliptical process} if its finite-dimensional marginals are elliptical with the same radial density $\varphi$, that is $(\mathbf{v}_i(t_1), \dots, \mathbf{v}_i(t_k)) \sim Ell(\mathbf{a}(t_1,\dots, t_k), \mathbf{S}(t_1,\dots, t_k), \varphi)$, and $\varphi$ is strictly decreasing. Here, $\mathbf{a}$ denotes the shift vector and $\mathbf{S}$ the structural matrix.

\begin{corollary}[Optimality of $DDk$- and $DD\alpha$-rules]\label{cor2}
 Assume that the processes $\mathfrak{G}_0$ and $\mathfrak{G}_1$ are strictly unimodal elliptical and have the same radial density, and choose $L^*$ and $S^*$ as above. Then

 (a) the following rules operating on $LS$-transformed data are Bayes optimal, relative to the
${\cal T}$-marginals:
\begin{description}
  \item[(i)] $DDk$-$M$,
    \item[(ii)] $DDk$-$S$,
      \item[(iii)] $DDk$-$P$,
\end{description}
      as $m,n,k\to\infty$  with $\frac km \to 0$ and $\frac kn \to 0$.

(b) If, in addition, the structural matrices coincide, $\mathbf{S}_0({\cal T})=\mathbf{S}_1({\cal T})$, and
the priors of class membership are equal, then the following rules operating on $LS$-transformed data are Bayes optimal, relative to the
${\cal T}$-marginals:
\begin{description}
  \item[(iv)] the maximum-depth rule,
  \item[(v)] the $DD\alpha$-$M$ rule,
  \item[(vi)] the $DD\alpha$-$S$ rule,
  \item[(vii)] the $DD\alpha$-$P$ rule,
  \end{description}
as $m,n\to\infty$.
  \end{corollary}

   Proof: Parts (i) to (iii) follow from Proposition \ref{prop1} above and Theorem 3.5 in~\cite{Vencalek11}.
Parts (v) to (vii) of the Corollary \ref{cor2} are a consequence of
Proposition \ref{prop1} and Proposition \ref{Prop2}.  Part (iv) is deduced from Proposition \ref{prop1} and \cite{GhoshC05b}, who demonstrate that
the maximum-depth rule is asymptotically equivalent to the Bayes rule if the two distributions
have the same prior probabilities and are elliptical in $\mathbb{R}^\ell$ with only a location difference.
\hfill $\Box$

The above optimality results regard rules that operate on a {\it DD}-plot of $LS$-transformed data.
For comparison, we will also apply the $kNN$-rule directly to $LS$-transformed data.

\begin{corollary}[Optimality of $kNN$]\label{cor5}
The $kNN$-classifier applied to $LS$-transformed data (where $L^*$ and $S^*$ are selected as above)
is Bayes optimal, relative to the ${\cal T}$-marginals.
\end{corollary}
Proof: This follows from Proposition \ref{prop1} and the universal consistency of the $kNN$ rule; see
\cite{DevroyeGL96}, Ch.\ 11. \hfill $\Box$

\subsection{Functional optimality}\label{ssec:funopt}

Next let us consider the asymptotic behavior of our procedure regarding the functions themselves.
The above results refer to Bayes optimality with respect to the $\ell$-dimensional marginal distributions of the two processes that correspond to a fixed set $\mathcal{T}$ of $\ell$ discretization points. Obviously they allow no inference on the whole functions.

In order to classify asymptotically with respect to the processes, we introduce a \emph{sequential sampling scheme} that evaluates the functions at an increasing number of discretization points. For this, a slightly different notation will be needed.

For each $k\in \mathbb{N}$ consider a set of discretization points,
${\cal T}_k=\{t_{k,1}, t_{k,2},\dots, t_{k,\ell_k}\}$, such that  ${\cal T}_k \subset {\cal T}_{k+1}$, $t_{k,0}=0$, $t_{k,\ell_k}=T$, $t_{k,j+1}>t_{k,j}$ for all $j$, and
\begin{equation}\label{condition D}
\max_{j=0,\dots,\ell_k} |t_{k,j+1}- t_{k,j}| \to 0 \quad {\rm when}\; k\to \infty\,.
\end{equation}
For $k\in \mathbb{N}$ assume that $\mathbf{\tilde v}_{1},\mathbf{\tilde v}_{2}, \dots ,\mathbf{\tilde v}_{k}$ are i.i.d.\ sampled paths from $\mathbf{\tilde v}\sim \mathfrak{G}_0$, and that each $\mathbf{\tilde v}_{j}$ is observed at times $t_{k,1},\dots, t_{k, \ell_k}$, $j=1,2,\dots, k$.
Similarly assume that $\mathbf{\tilde w}_{1},\mathbf{\tilde w}_{2}, \dots ,\mathbf{\tilde w}_{k}, \dots$ are i.i.d.\ from $\mathbf{\tilde w}\sim \mathfrak{G}_1$, each $\mathbf{\tilde w}_{j}$ being observed at times $t_{k,1},\dots, t_{k, \ell_k}$, and that the two samples are independent of each other.

In this setting, we apply our classification procedure with $L$ and $S$ depending on $k$.
By cross-validation we select an optimal $L_{k}$ from the set
\[ {\cal M}_{k}=\left\{ L\in \mathbb{N} | L\le \frac T{\min_j(t_{k,j+1}-t_{k,j})} + 1 \right\}.
\]
For $i=1,2,\dots k$, let $\mathbf{v}^k_i$ be the linearly interpolated function that is supported at  $\bigl(t_{k,1},\mathbf{\tilde v}_i(t_{k,1})\bigr)$, $\dots$, $\bigl(t_{k,\ell_k},\mathbf{\tilde v}_i(t_{k,\ell_k})\bigr)$, and define $\mathbf{w}^k_i$ similarly.
 $\mathfrak{G}_{0,k}$ denotes the empirical distribution on $\mathbf{v}^k_1,\dots, \mathbf{v}^k_k$, and and $\mathfrak{G}_{1,k}$ that on $\mathbf{w}^k_1,\dots, \mathbf{w}^k_k$.
Then, obviously, the classification based on the $LS$-transformation, with $L_k$ being selected from ${\cal M}_{k}$ via cross validation, is optimal regarding the empirical distributions $\mathfrak{G}_{0,k}$ and $\mathfrak{G}_{1,k}$.


Now we can extend Corollary 1 to the case of process classification:

\begin{theorem}[Bayes optimality with respect to Gaussian processes]
 \label{cor2a}
 Let $\mathfrak{G}_0$ and $\mathfrak{G}_1$ be Gaussian processes that have piecewise smooth paths, and let the priors of class membership be equal. Under the above sequential sampling scheme and if  $L_k$ is selected from ${\cal M}_{k}$, the following rules operating on the $LS$-transformed data are almost surely Bayes optimal regarding the processes $\mathfrak{G}_{0}$ and $\mathfrak{G}_{1}$:
\begin{description}
  \item[(i)] quadratic discriminant analysis (QDA),
  \item[(ii)] linear discriminant analysis (LDA), if $\mathfrak{G}_0$ and $\mathfrak{G}_1$ have the same covariance function.
\end{description}
\end{theorem}

Proof: As the processes are Gaussian, we obtain that
\begin{align}\label{Sigmas1}
\bigl(\mathbf{v}_i^k(t_{k,1}), \dots, \mathbf{v}_i^k(t_{k,\ell_k})\bigr) \sim (i.i.d.) N(\mathbf{\mu}_{0,k}, \mathbf{\Sigma}_{0,k})\,,\quad i=1,\dots, k\,,\\
\bigl(\mathbf{w}_i^k(t_{k,1}), \dots, \mathbf{w}_i^k(t_{k,\ell_k})\bigr) \sim (i.i.d.) N(\mathbf{\mu}_{1,k}, \mathbf{\Sigma}_{1,k})\,,\quad i=1,\dots, k\,,
\end{align}
where $\mu_{j,k}\in \mathbb{R}^{\ell_k}$, and the $\mathbf{\Sigma}_{j,k}$ are $(\ell_k \times \ell_k)$-covariance matrices, $j=0,1$.
 Observe that the optimal (i.e. crossvalidated) error rate converges to the probability of error.
 If $L_k$ is selected from ${\cal M}_{k}$, then, as in Corollary 1, the Bayes risk is achieved with respect to the empirical distributions.
Theorem 1(ii) of \cite{NagyGH15} says that under the above sampling scheme it holds
\begin{equation} \label{weakconv}
P[\mathfrak{G}_{0,k} \to \mathfrak{G}_{0}\ {\rm weakly}]=1 \quad {\rm and } \;\; P[\mathfrak{G}_{1,k} \to \mathfrak{G}_{1}\ {\rm weakly}]=1.
\end{equation}
As for every $k$ the Bayes risk is achieved with respect to $\mathfrak{G}_{0,k}$ and $\mathfrak{G}_{1,k}$, (\ref{weakconv}) implies that, in the limit ($k\to \infty$), the Bayes risk is almost surely achieved with respect to the processes $\mathfrak{G}_{0}$  and $\mathfrak{G}_{1}$.
\hfill $\Box$

\section{Choosing the dimensions $L$ and $S$}\label{sec:lss}
Clearly, the performance of our classifier depends on the choice of $L$ and $S$, which has still to be discussed. In what follows we assume that the data is given as functional values at a (usually large) number of discretization points. Let $M$ denote the number of these points.

\cite{DelaigleHB12} propose to perform the classification in a finite-dimensional space that is based on a subset of discretization points selected to minimize the average error.
But these authors do not offer a construction rule for the task but rely on multi-layer leave-one-out cross-validation, which is very time-consuming.
Having recognized this problem they suggest some time-saving modifications of the cross-validation procedure.
Clearly, the computational load is then determined by the cross-validation scheme used. It naturally depends on the size of the data sample, factored with the training time of the finite-dimensional classifier, which may also include parameters to be cross-validated.

The  \cite{DelaigleHB12} approach (abbreviated \emph{crossDHB} for short) suggests a straightforward procedure for our problem of constructing an $LS$-space:
allow for a rich initial set of possible pairs $(L,S)$ and then use cross-validation in selecting a pair that (on an average) performs best.
The initial set shall consist of all pairs $(L,S)$, say, satisfying $2\le L+S\le M/2$.
Other upper bounds may be used, e.g.\ if $M/2$ exceeds the number of observations. In addition, a dimension-reducing technique like PCA or factor analysis may be in place.
But this cross-validation approach (\emph{crossLS} for short), similar to the one of~\cite{DelaigleHB12}, is still time-consuming; see Section~\ref{ssec:time} for computing times.
The problem of selecting an appropriate pair $(L^\star,S^\star)$ remains challenging.

In deciding about $L$ and $S$, we will consider the observed performance of the classifier as well as some kind of worst-case performance.
Fortunately, the conservative error bound of Vapnik and Chervonenkis \citep{VapnikC74}, see also \cite{DevroyeGL96},
provides some guidance.
The main idea is to measure how good a certain location-slope space can be at all, by referring to the \emph{worst-case empirical risk of a linear classifier} on the training classes.
Clearly, the class of linear rules is limited, but may be regarded as an approximation.
Also, this limitation keeps the deviation $\Delta\epsilon$ from empirical risk small and allows for its implicit comparison with the empirical risk $\epsilon$ itself. Moreover, in choosing the dimension of the location-slope space we may adjust for the required complexity so that finally the separation rule is close to a linear rule.

Here, \textit{linear discriminant analysis} (LDA) is used for the linear classification. Although other approaches like perceptron,
regression depth, support vector machine,
or $\alpha$-procedure
can be employed instead, we use LDA as it is fast and linear.
Below we derive the Vapnik-Chervonenkis conservative bound for our particular case; this is similar to Theorem 5.1 in~\cite{VapnikC74}. Here we use another bound on the cardinality of the class of separating rules \citep[see also][]{DevroyeGL96}.

 Consider the $LS$-transformed data $Y_0$ and $Y_1  \subset \mathbb{R}^{L+S}$, as two samples of random vectors that are distributed as $\mathfrak{G}_0^{L+S}$ and $\mathfrak{G}_1^{L+S}$,
 respectively. Let $\mathcal{L}$ be a class of separation rules ${\boldsymbol r}:\mathbb{R}^{L+S}\mapsto\{0,1\}$, where two different rules separate $Y_0\cup Y_1$ into two different subsets. Then each ${\boldsymbol r}$ produces an expected classification error,
\begin{eqnarray}
    {\mathcal E}({\boldsymbol r},\mathfrak{G}_0^{L+S},\mathfrak{G}_1^{L+S}) & = & \pi_0\int_{\mathbb{R}^{L+S}}I\bigl({\boldsymbol r}({\mathbf y})=1\bigr)d\mathfrak{G}_0^{L+S}(\mathbf{y})\\
    & + & \pi_1\int_{\mathbb{R}^{L+S}}I\bigl({\boldsymbol r}({\mathbf y})=0\bigr)d\mathfrak{G}_0^{L+S}(\mathbf{y}),\nonumber
\end{eqnarray}
with $\pi_0$ and $\pi_1$ being prior probabilities of $\mathfrak{G}_0^{L+S}$ and $\mathfrak{G}_1^{L+S}$, and $I(\cdot)$ being an indicator function.
 As mentioned above, ${\mathcal E}({\boldsymbol r},\mathfrak{G}_0^{L+S},\mathfrak{G}_1^{L+S})$ can be consistently estimated by leave-one-out cross-validation from $Y_0$ and  $Y_1$. However, this estimate comes at substantial computational cost. On the other hand, based on the empirical risk $\epsilon$,
 the expectation ${\mathcal E}({\boldsymbol r},\mathfrak{G}_0^{L+S},\mathfrak{G}_1^{L+S})$ can be bounded from above. The empirical risk is given by
\begin{equation}
    \epsilon({\boldsymbol r},Y_0,Y_1)=\frac{1}{m+n}\Bigl(\sum_{i=1}^m I\bigl({\boldsymbol r}({\mathbf y}_i)=1\bigr)+\sum_{i=m+1}^{m+n}I\bigl({\boldsymbol r}({\mathbf y}_i)=0\bigr)\Bigr).
\end{equation}
As discussed before, we constrain $\mathcal{L}$ to be the class of linear rules. Then the number of possible rules equals the number $\#\mathcal{L}$
of different separations of $Y_0\cup Y_1$ by hyperplanes $\subset\mathbb{R}^{L+S}$. Note that $C(N,d)=2\sum_{k=0}^{d-1}{N - 1 \choose k}$ is the number of possible different separations of $N$ points in ${\mathbb R}^d$ that can be achieved by hyperplanes containing the origin. Consequently, it holds
$\#\mathcal{L}\le C(m + n, L + S + 1)$, and equality is attained if $Y_0\cup Y_1$ are in general position. Applying Hoeffding's~(\citeyear{Hoeffding63}) inequality, the probability that the highest (over the class $\mathcal{L}$) excess of the classification error over the empirical risk will not exceed some positive value $\Delta\epsilon$ can be bounded from above:
\begin{eqnarray}\label{eqn:Hoeffding}
    && P\bigl(\sup_{\boldsymbol{r}\in\mathcal{L}}\{{\mathcal E}({\boldsymbol r},\mathfrak{G}_0^{L+S},\mathfrak{G}_1^{L+S})-\epsilon({\boldsymbol r},Y_0,Y_1)\}>\Delta\epsilon\bigr)\\
    && \le C(m + n, L + S + 1)e^{-2(m+n)(\Delta\epsilon)^2}.\nonumber
\end{eqnarray}
This allows to formulate the following proposition:

\begin{figure*}[t!]
    \centering
	\includegraphics[keepaspectratio=true,scale=0.415]{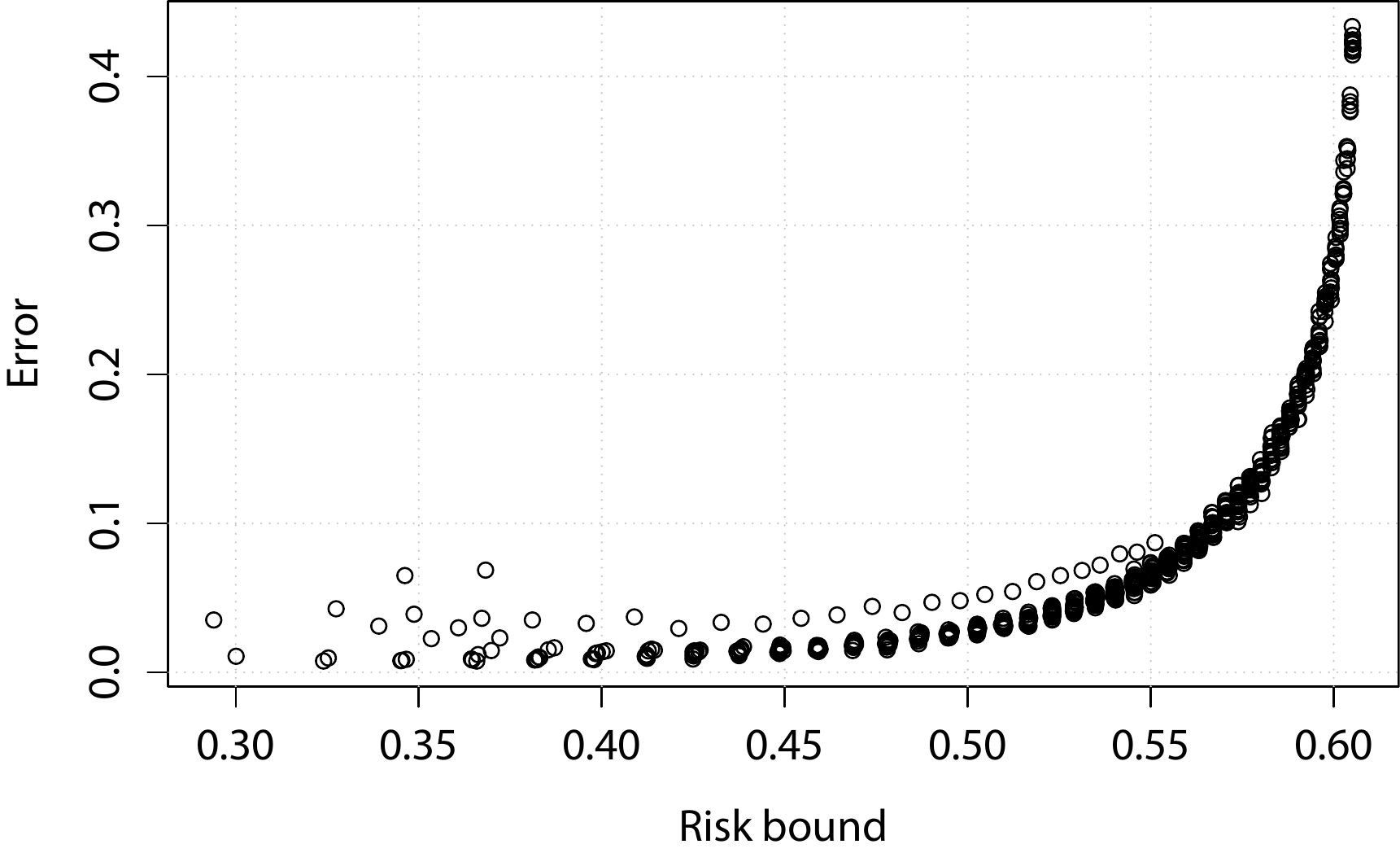}
    \includegraphics[keepaspectratio=true,scale=0.415]{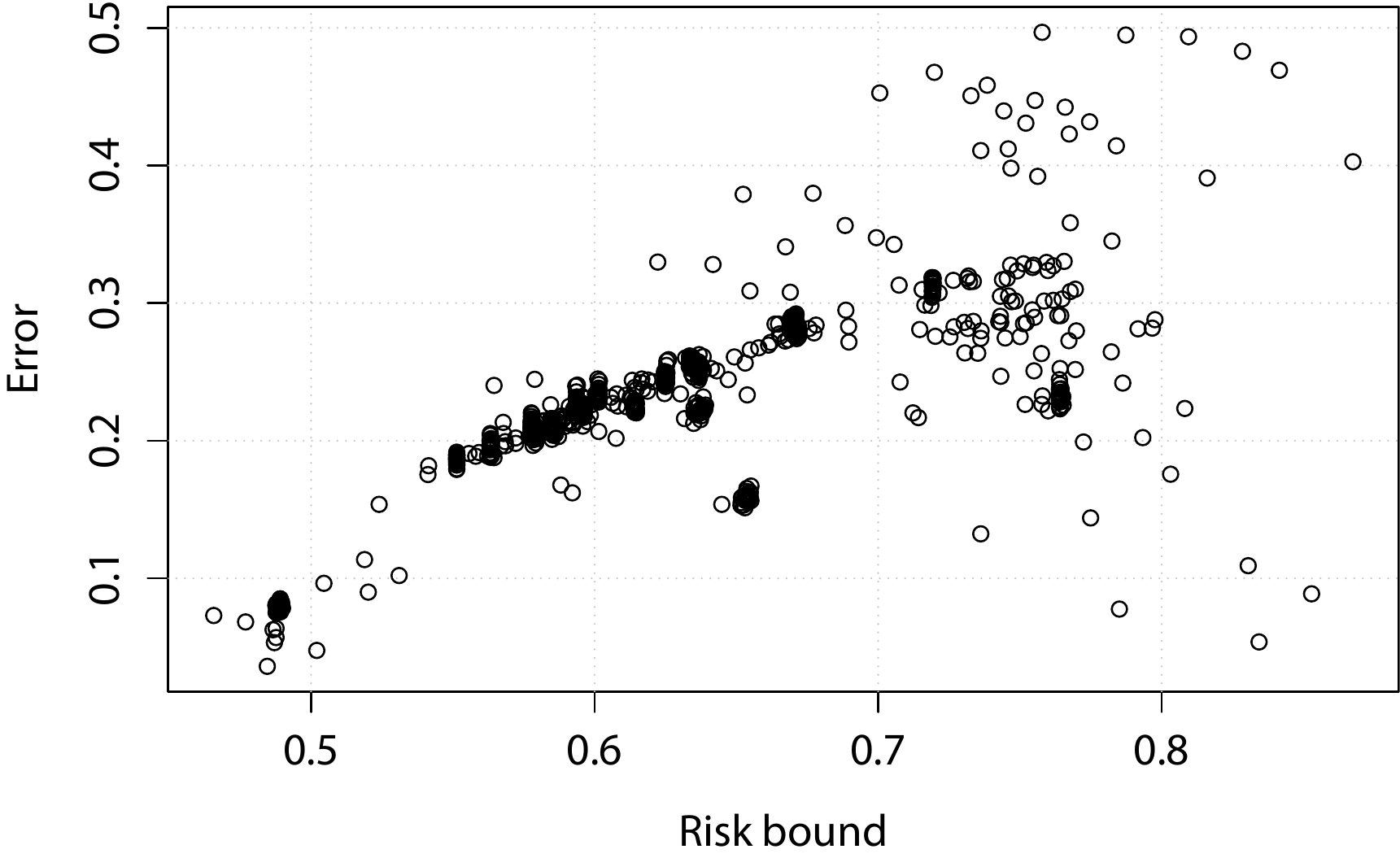}
    \caption{Average misclassification error (ordinate) to maximal risk (abscissa) of the $DD\alpha$-classifier based on Mahalanobis depth for Model 1 (left) and Model 2 (right) of \cite{CuevasFF07}.}
    \label{fig:VapnikC1}
\end{figure*}

\begin{figure*}[t!]
    \centering
	\includegraphics[keepaspectratio=true,scale=0.415]{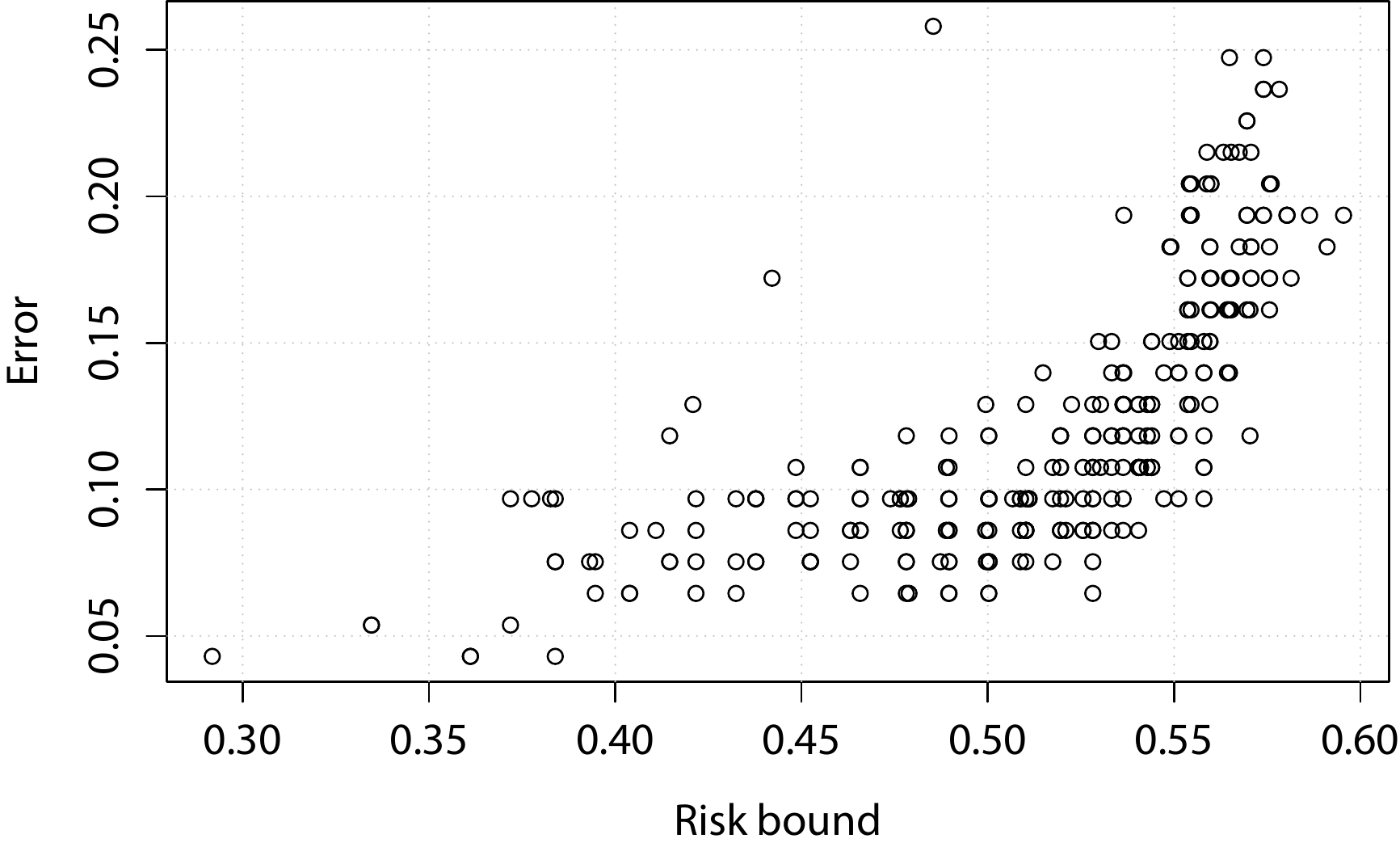}
    \includegraphics[keepaspectratio=true,scale=0.415]{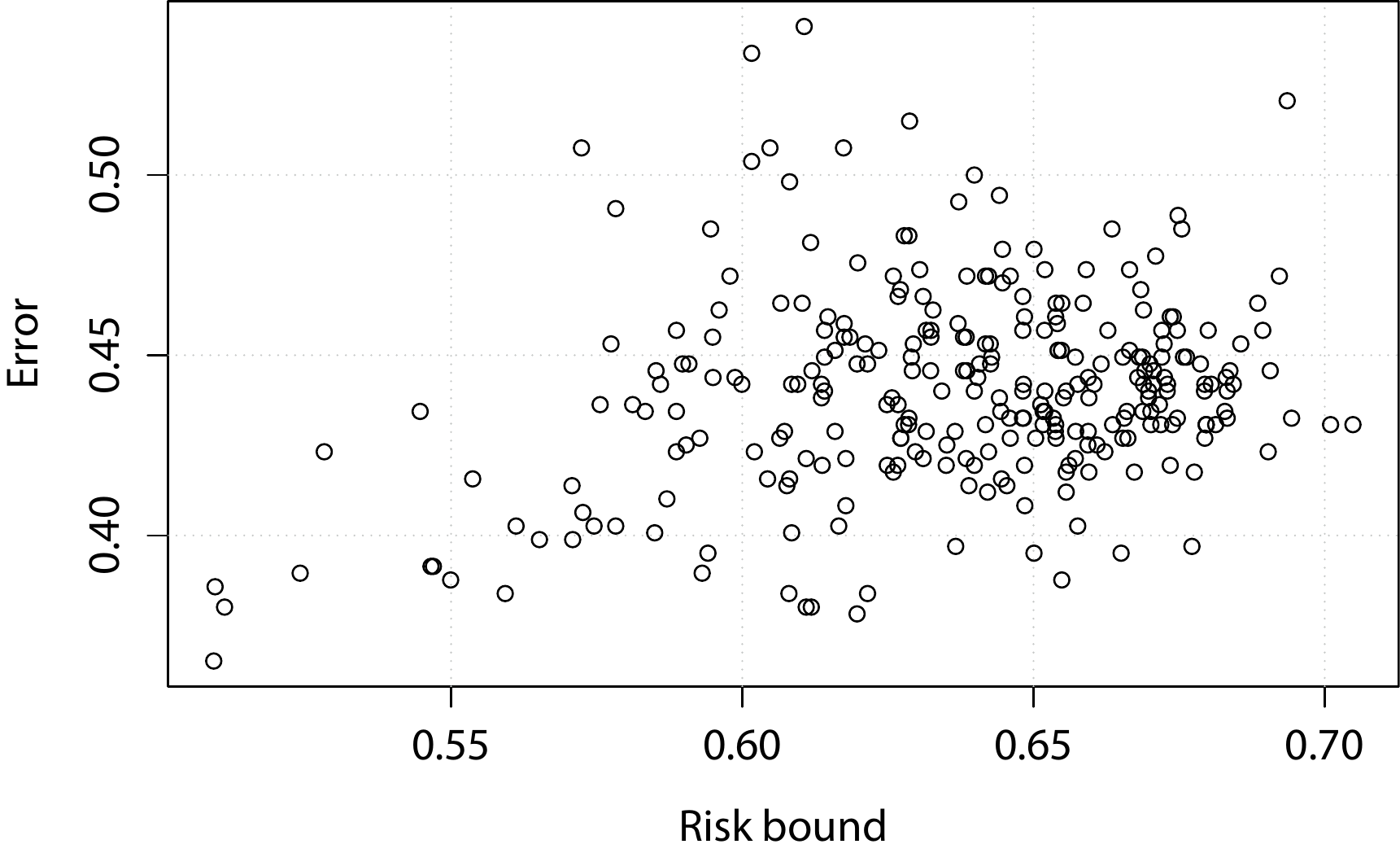}
	\caption{Average misclassification rate (ordinate) to maximal risk (abscissa) of the $DD\alpha$-classifier based on Mahalanobis depth for \emph{growth  data} (left) and \emph{medflies data} (right).}

    \label{fig:VapnikC2}
\end{figure*}

\begin{proposition}\label{Prop3}
    For the class of linear separating rules in $\mathbb{R}^{L+S}$, it holds with probability $1-\eta$:
    \begin{equation}\label{equ:Vapnik}
    {\mathcal E}({\boldsymbol r},\mathfrak{G}_0^{L+S},\mathfrak{G}_1^{L+S})\le\epsilon({\boldsymbol r},Y_0,Y_1)+\sqrt{\frac{\ln{C(m + n, L + S + 1)} - \ln{\eta}}{2(m + n)}}.
\end{equation}
\end{proposition}
Proof: Requiring $P\bigl(\sup_{\boldsymbol{r}\in\mathcal{L}}\{{\mathcal E}({\boldsymbol r},\mathfrak{G}_0^{L+S},\mathfrak{G}_1^{L+S})-\epsilon({\boldsymbol r},Y_0,Y_1)\}>\Delta\epsilon\bigr)\le\eta$ yields $\eta=C(m + n, L + S + 1)e^{-2(m+n)(\Delta\epsilon)^2}$. Solving the last w.r.t. $\Delta\epsilon$ gives $\Delta\epsilon=\sqrt{\frac{\ln{C(m + n, L + S + 1)} - \ln{\eta}}{2(m + n)}}$. Then, with probability $1-\eta$, equation (\ref{equ:Vapnik}) follows immediately.
\hfill $\Box$

The probability $\eta$ is fixed to some small constant. Here we use $\eta=\frac{1}{m + n}$, given the sample size. The goodness of a certain choice of $L$ and $S$ is then measured by
\begin{equation}\label{eqn:emax}
    \epsilon_{max}=\epsilon+\sqrt{\frac{\ln{\bigl((m + n)C(m + n, L + S + 1)\bigr)}}{2(m + n)}}.
\end{equation}
We restrict our choice of $L$ and $S$ to those that make $\epsilon_{max}$ small. An intuitive justification
is the following: $\epsilon$ refers to empirical risk, while the second term penalizes the dimension, which balances fit and complexity.

To find out whether the proposed bound really helps in finding proper dimensions $L$ and $S$, we first apply our approach to two simulated data settings of \cite{CuevasFF07}, called `Model\,1' and `Model\,2' (both $M=51$). The data generating process of \cite{CuevasFF07} is described in Section~\ref{ssec:simresults} below. We determine $L$ and $S$, use the Mahalanobis depth to construct a $DD$-plot and apply the $DD\alpha$-classifier, which is abbreviated in the sequel as $DD\alpha$-$M$.
For each pair $(L,S)$ with $L+S\ge2$ and $L+S\le26$, the risk bound $\epsilon_{max}$ and the average classification error (ACE) are calculated by averaging over 100 takes and plotted in
Figure~\ref{fig:VapnikC1}. Note that these patterns look similar when spatial or projection depth is used. Further, for the two benchmark data problems, we estimate ACE by means of leave-one-out cross-validation (see Figure~\ref{fig:VapnikC2}). Here all $(L,S)$-pairs are considered with $L+S\ge2$ and $L+S\le16$.

As expected, Figures~\ref{fig:VapnikC1} and~\ref{fig:VapnikC2} largely support the statement ``the less $\epsilon_{max}$ the smaller the ACE''.
Although for the challenging \emph{medflies data} (Figure~\ref{fig:VapnikC2}, right) the plot remains a fuzzy scatter, {a substantial part of points with small error have low values of $\epsilon_{max}$. (Even more, the $(L,S)$-pair with the smallest $\epsilon_{max}$ actually corresponds to the smallest error, see the point in the lower left corner.) Thus $\epsilon_{max}$} still guides us in configuring the location-slope space, as it is affirmed by our experimental results in Section~\ref{ssec:benchmarks} below. Computing $\epsilon_{max}$ involves a single calculation of the LDA-classifier (\emph{viz.} to estimate $\epsilon$), which is done in negligible time. Then,
the $(L,S)$-pair with the smallest $\epsilon_{max}$ can be picked. We restrict attention to a few $(L,S)$-pairs that deliver smallest values of $\epsilon_{max}$, and choose the best one by cross-validating over this tiny range. This new technique is employed here for space building.
We abbreviate it as \emph{VCcrossLS}.

\section{Experimental comparisons}\label{sec:experiments}
To evaluate the performance of the proposed methodology we compare it with several classifiers that operate
either on the original data or
in a location-slope space. After introducing those competitors (Section~\ref{ssec:competitors}) we present a simulation study (Section~\ref{ssec:simresults}) and a benchmark study (Section~\ref{ssec:benchmarks}), including a discussion of computation loads (Section~\ref{ssec:time}).
Implementation details of the experiments are provided in Appendix~A.

\subsection{Competitors}\label{ssec:competitors}
The new classifiers are compared with four classification rules: linear (LDA) and quadratic (QDA) discriminant analysis, $k$-nearest-neighbors ($kNN$) classification and the maximum-depth classification (employing the three depth notions),
all operating in a properly chosen location-slope space that is constructed with the bounding technique \emph{VCcrossLS} of Section~\ref{sec:lss}.
(One may argue that the $\epsilon_{max}$-based choice of $(L,S)$ is not generally suited for $kNN$, but it delivers comparable results in reasonable time, much faster than cross-validation over all $(L,S)$-pairs.)
Also, the four classifiers mentioned above, together with the two new ones (with all three depths), are used when the dimension of the location-slope space is determined by non-restricted cross-validation \emph{crossLS}.
For further comparison, all 12 classifiers are applied in the finite-dimensional space constructed according to the methodology \emph{crossDHB} of \cite{DelaigleHB12}.

{\em LDA} and {\em QDA} are calculated with classical moment estimates, and priors are estimated by the class portions in the training set.
We include $kNN$ in our competitors as it is Bayes-risk consistent in the finite-dimensional setting and generally performs very well in applications.
The {\em $kNN$-classifier} is applied to the location-slope data in its affine invariant form.
It is then defined as in (\ref{eqn:knn}), but with the Mahalanobis distance (determined from the pooled data) in place of the $L_\infty$-distance.
$k$ is selected by cross-validation.

As further competitors we consider three {\em maximum depth classifiers}. They are defined as
\begin{equation}
class({\mathbf x}) = \argmax_{i} \pi_i D(\mathbf{y}|Y_i)\,,
\end{equation}
with $D$ being either Mahalanobis depth $D^{Mah}$, or spatial depth $D^{Spt}$, or projection depth $D^{Prj}$. $\pi_i$ denotes the prior probability for class $i$. The priors are estimated by the class portions in the training set. This classifier is Bayes optimal if the data comes from an elliptical location-shift model with known priors.
For technical and implementation details the reader is referred to Appendix~A.

\subsection{Simulation settings and results}\label{ssec:simresults}
\begin{figure*}
    \centering
	\includegraphics[keepaspectratio=true,scale=0.415]{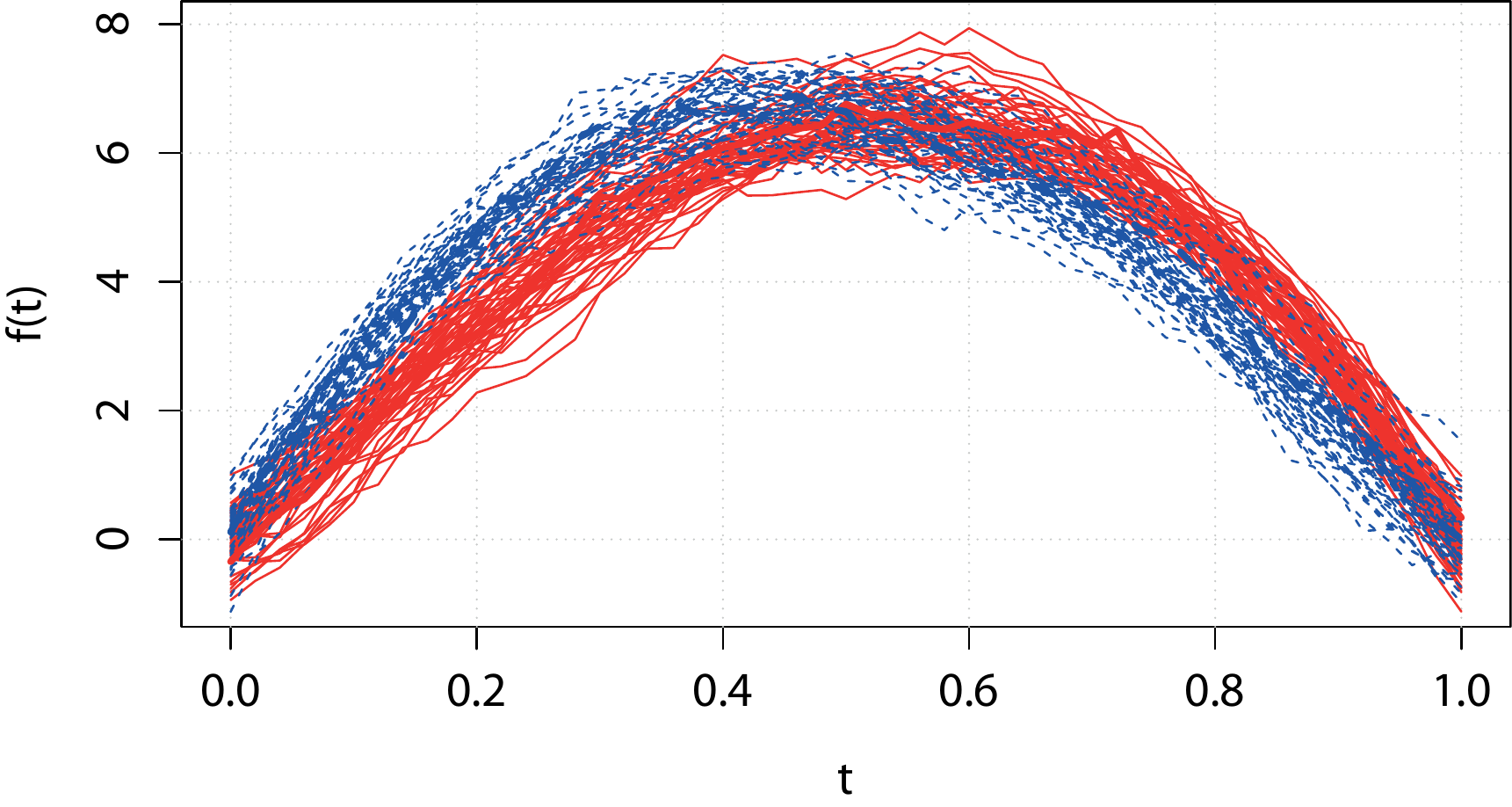}
    \includegraphics[keepaspectratio=true,scale=0.415]{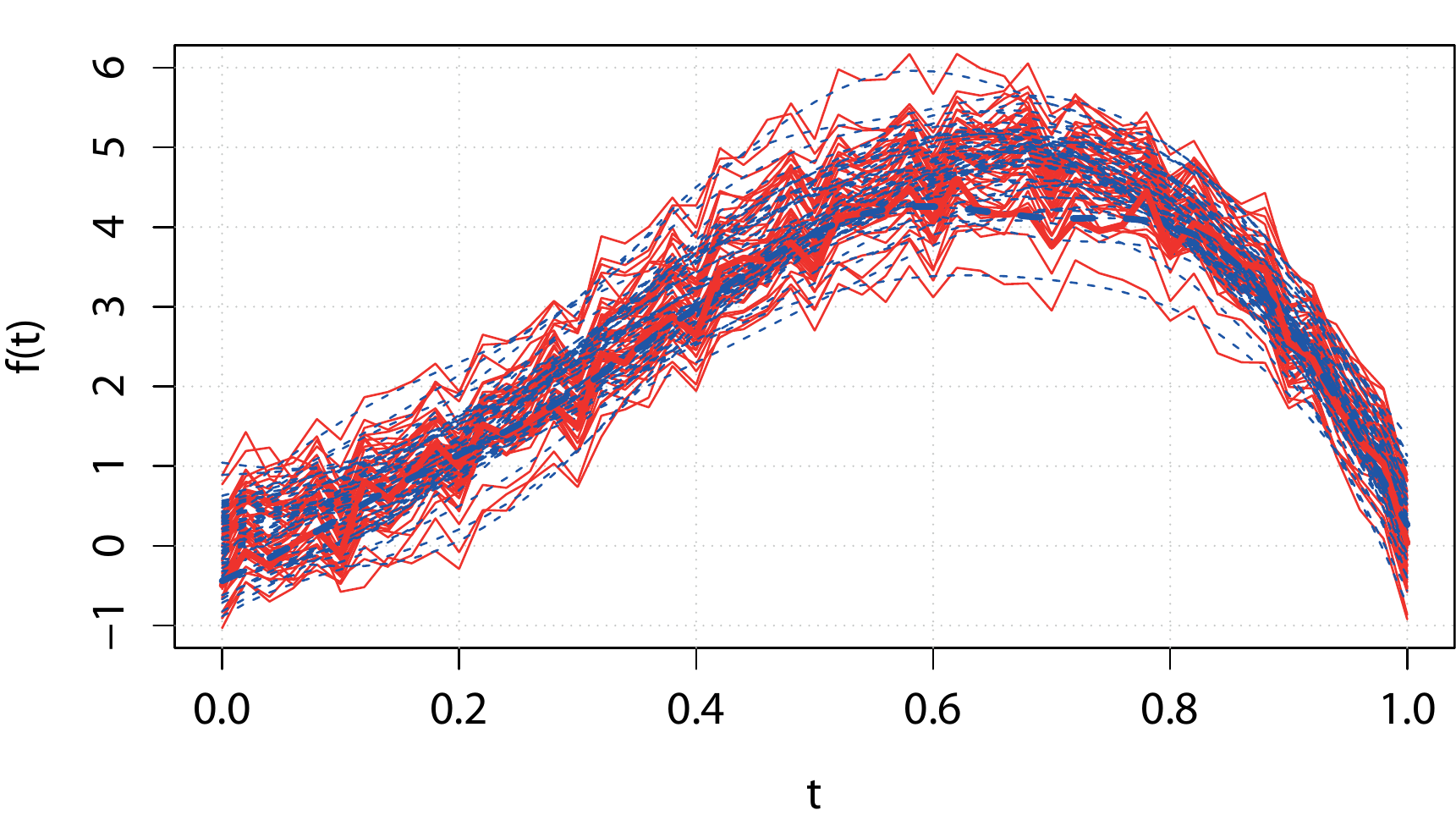}
    \caption{Synthetic data: Model 1 (left) and Model 2 (right) of \cite{CuevasFF07}.}
    \label{fig:models.CFF07}
\end{figure*}

Next we explore our methodology by applying it to the simulation setting of \cite{CuevasFF07}.
Their data are generated from two models, each having two classes. The first model is
{\em Model 1}:
\begin{eqnarray*}
  X_0&=&\{\mathbf{x}^0(t)|\mathbf{x}^0(t)=30(1-t)t^{1.2}+u(t)\}\,,\\
  X_1&=&\{\mathbf{x}^1(t)|\mathbf{x}^1(t)=30(1-t)^{1.2}t+u(t)\}\,,
\end{eqnarray*}
where $u(t)$ is a Gaussian process with $E[u(t)]=0$ and $Cov[u(s),u(t)]=0.2e^{-\frac{1}{0.3}|s-t|}$, discretized at 51 equally distant points on $[0,1]$ ($M=51$), see Figure~\ref{fig:models.CFF07} (left) for illustration. The functions are smooth and differ in mean only, which makes the classification task rather simple.

\begin{figure*}[!t]
    \centering
    \includegraphics[keepaspectratio=true,scale=0.445]{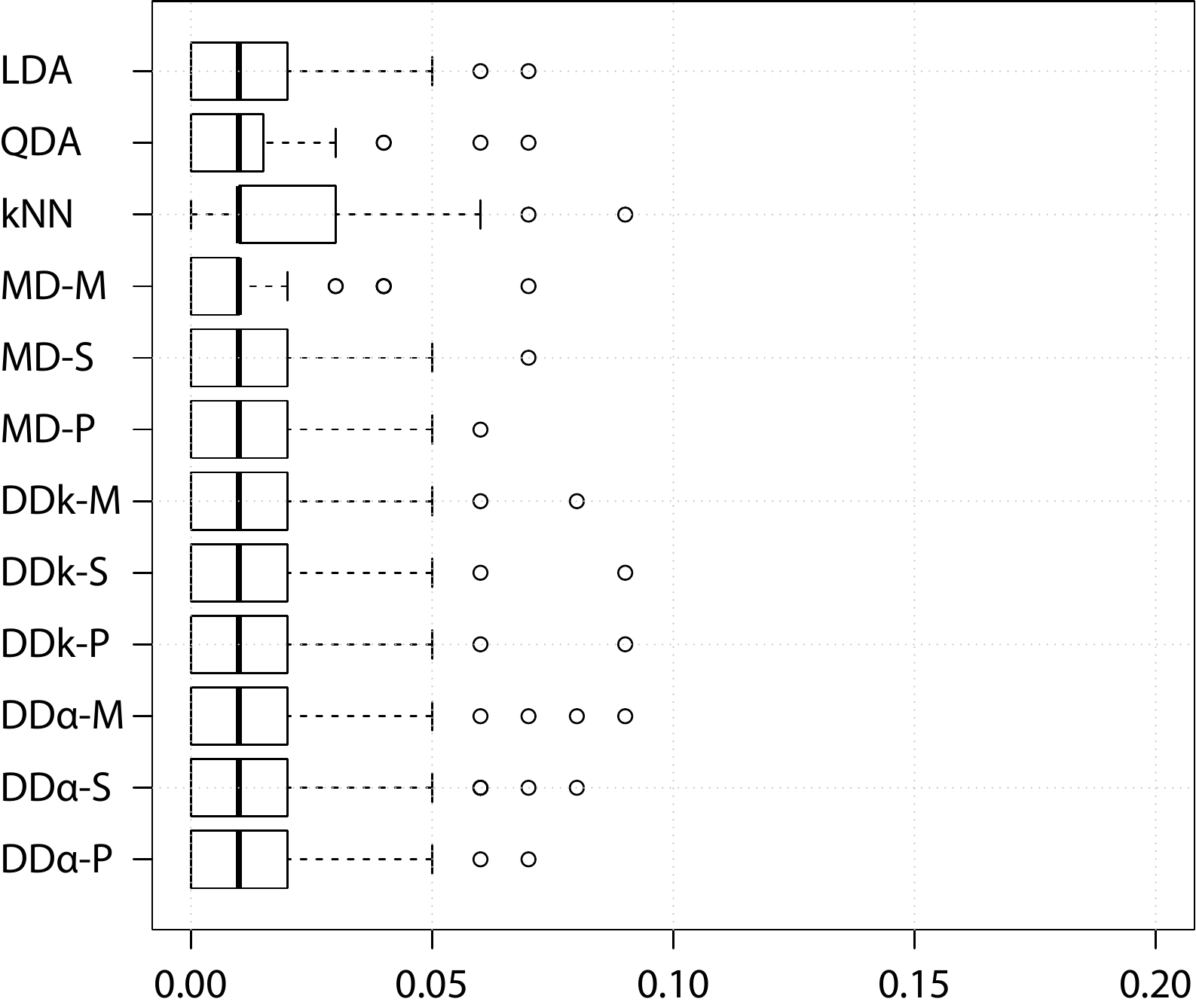}
    \includegraphics[keepaspectratio=true,scale=0.445]{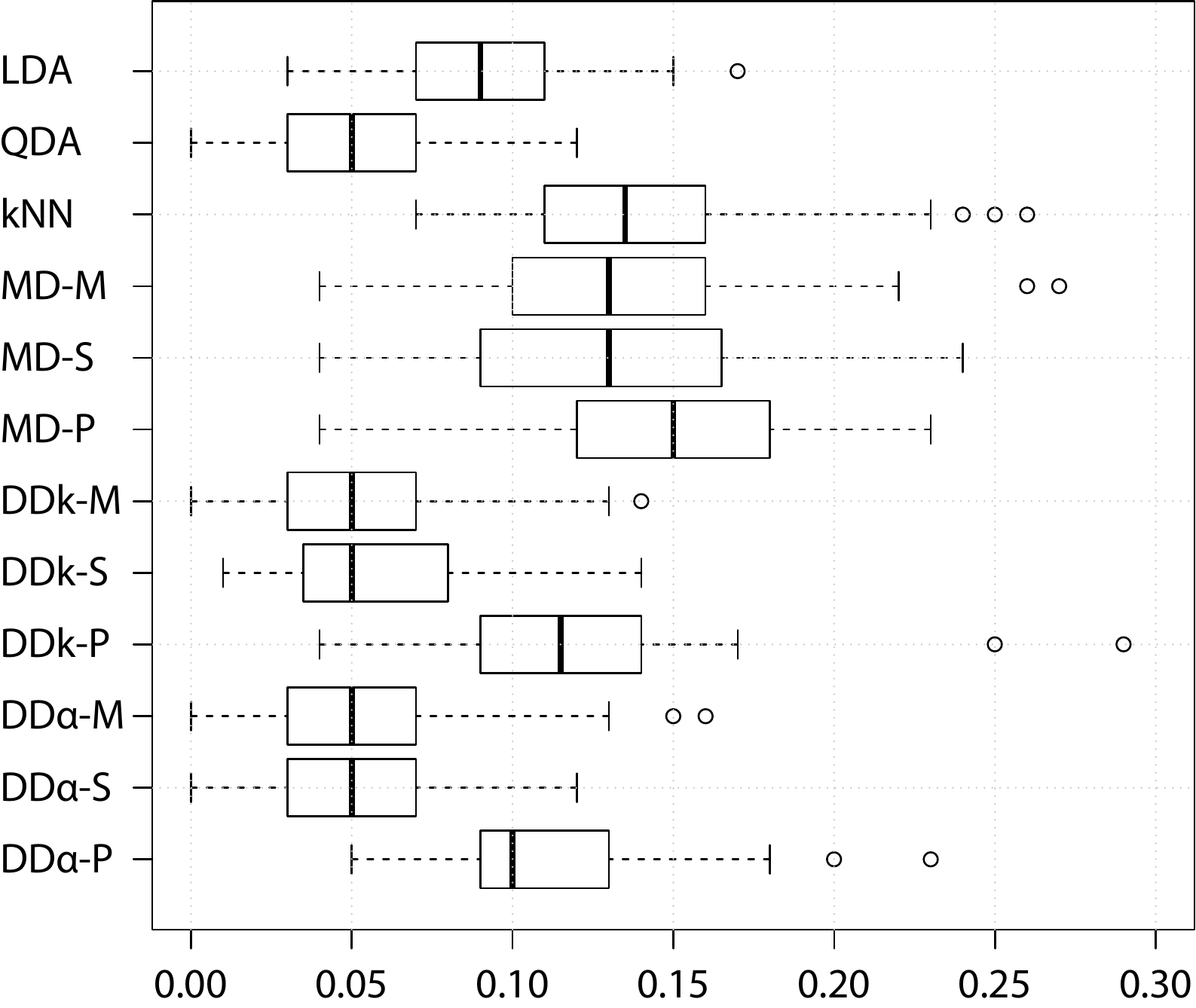}\\
    \indent\\
    \includegraphics[keepaspectratio=true,scale=0.445]{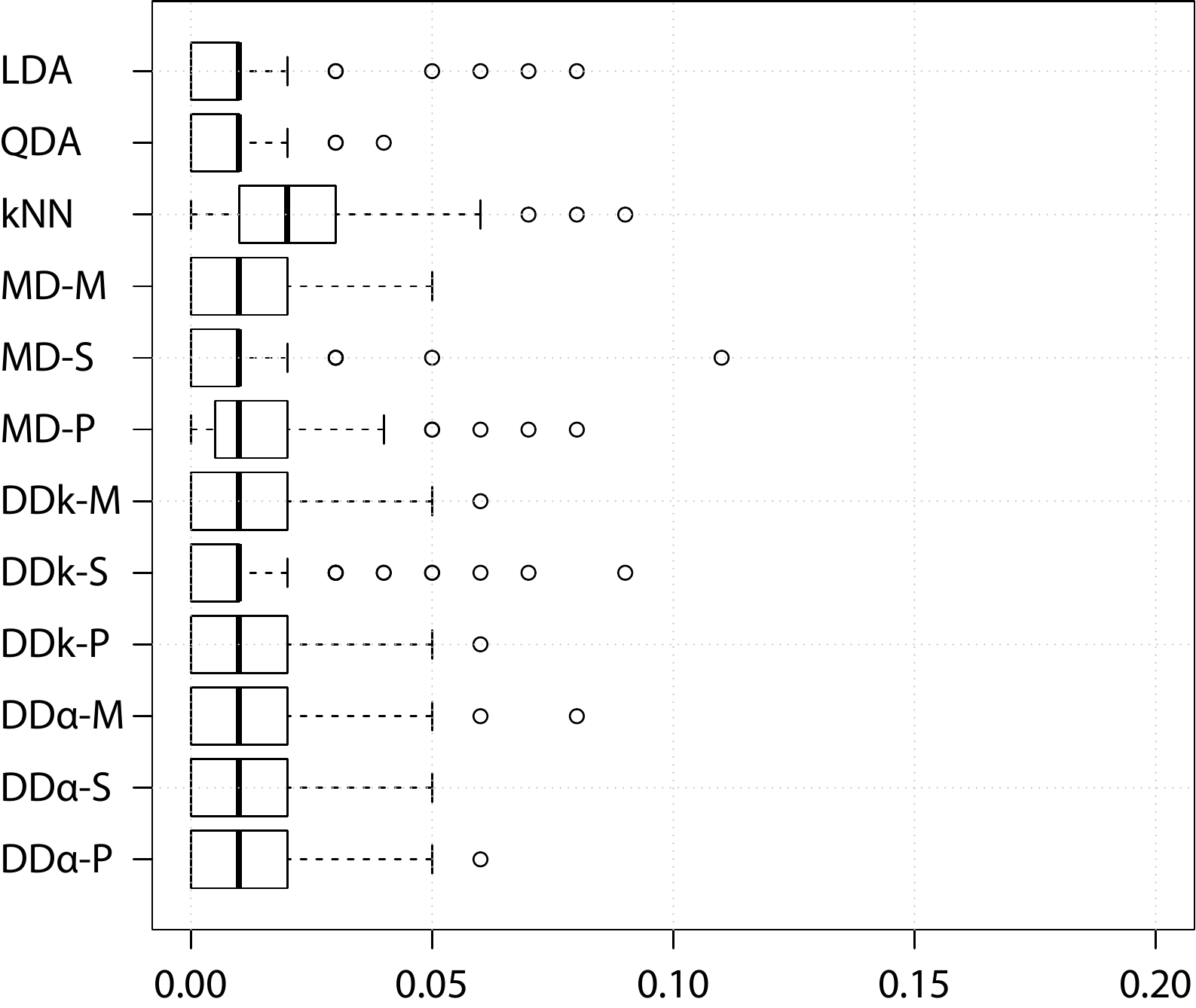}
    \includegraphics[keepaspectratio=true,scale=0.445]{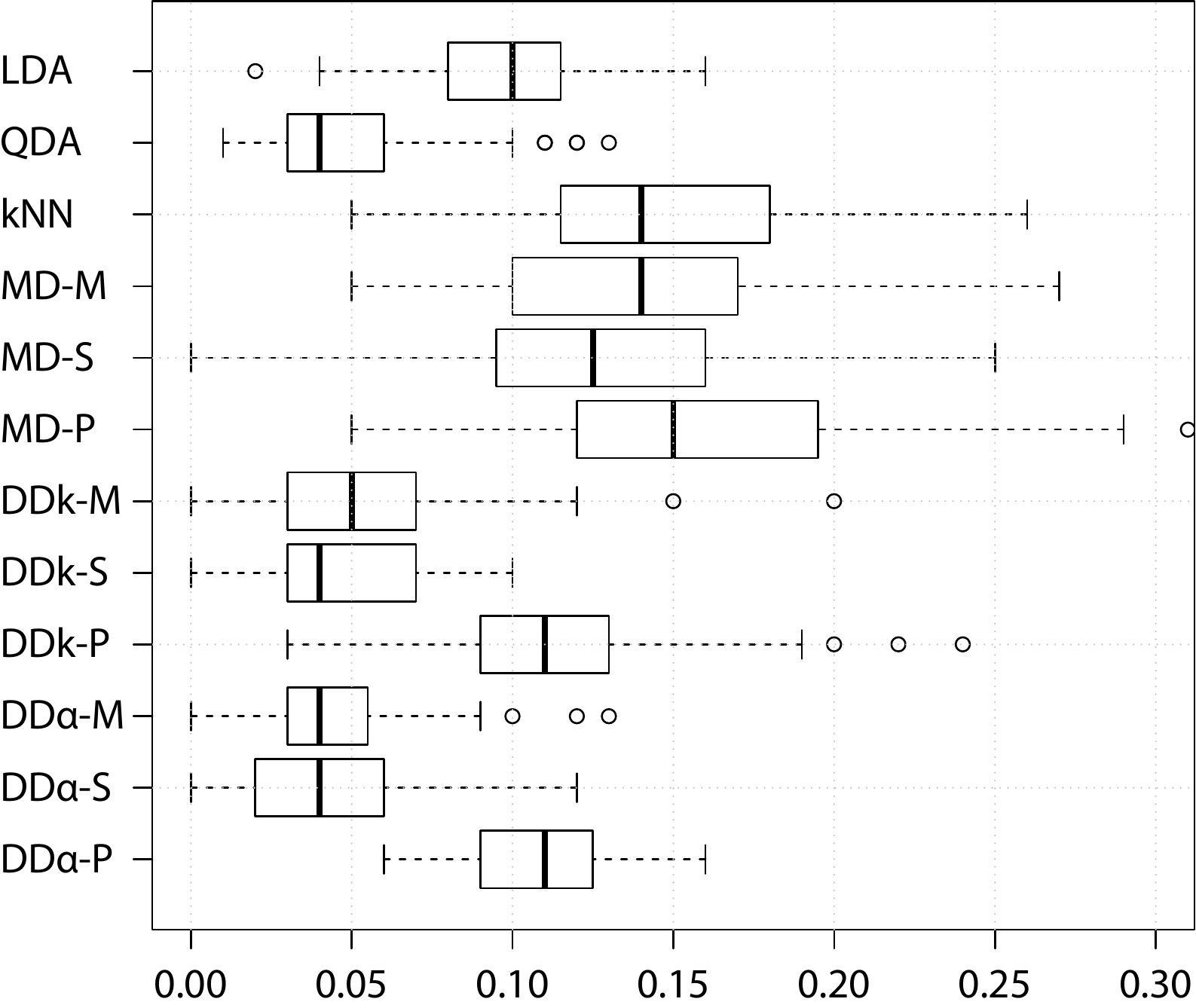}\\
    \indent\\
    \includegraphics[keepaspectratio=true,scale=0.445]{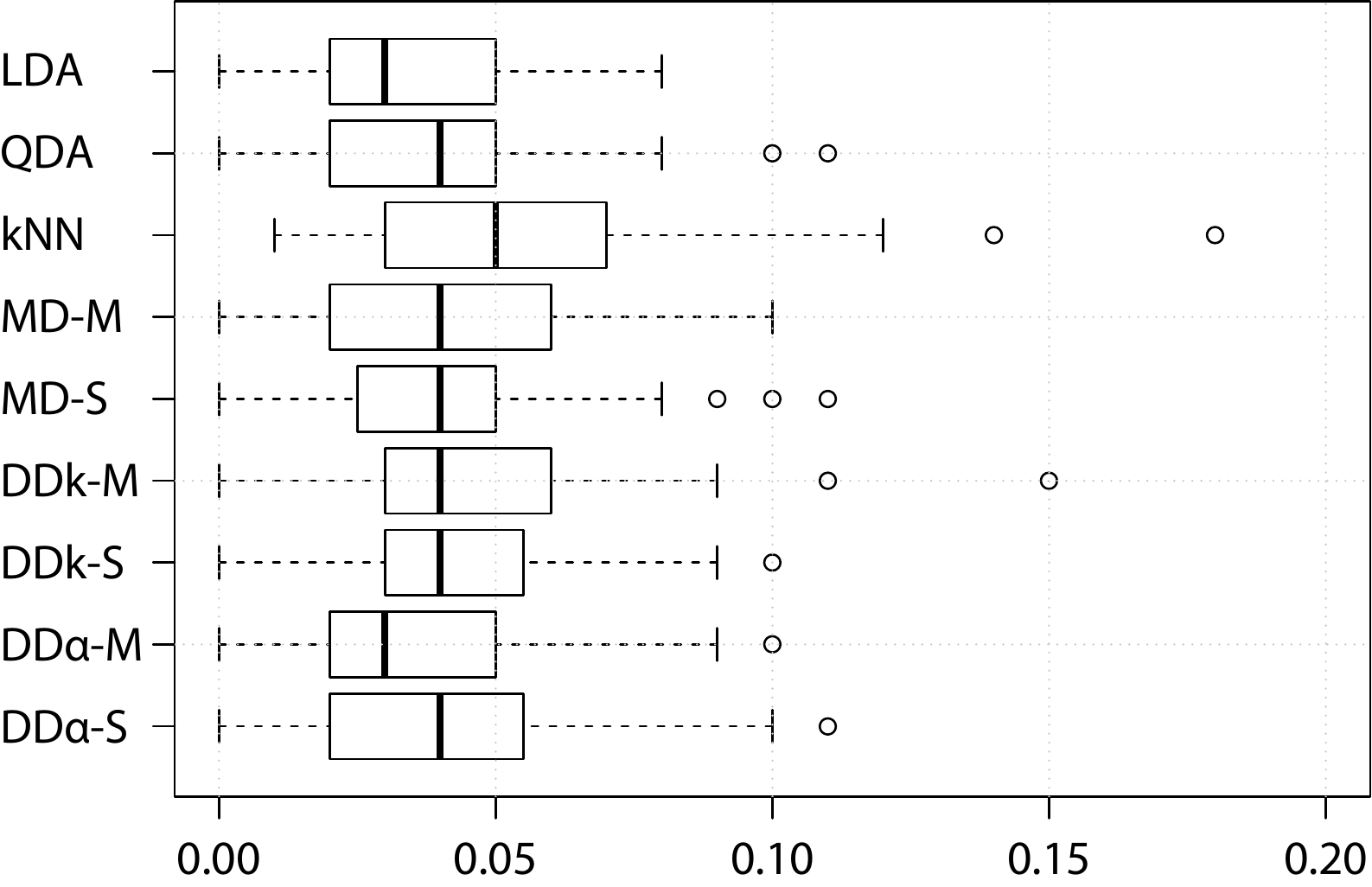}
    \includegraphics[keepaspectratio=true,scale=0.445]{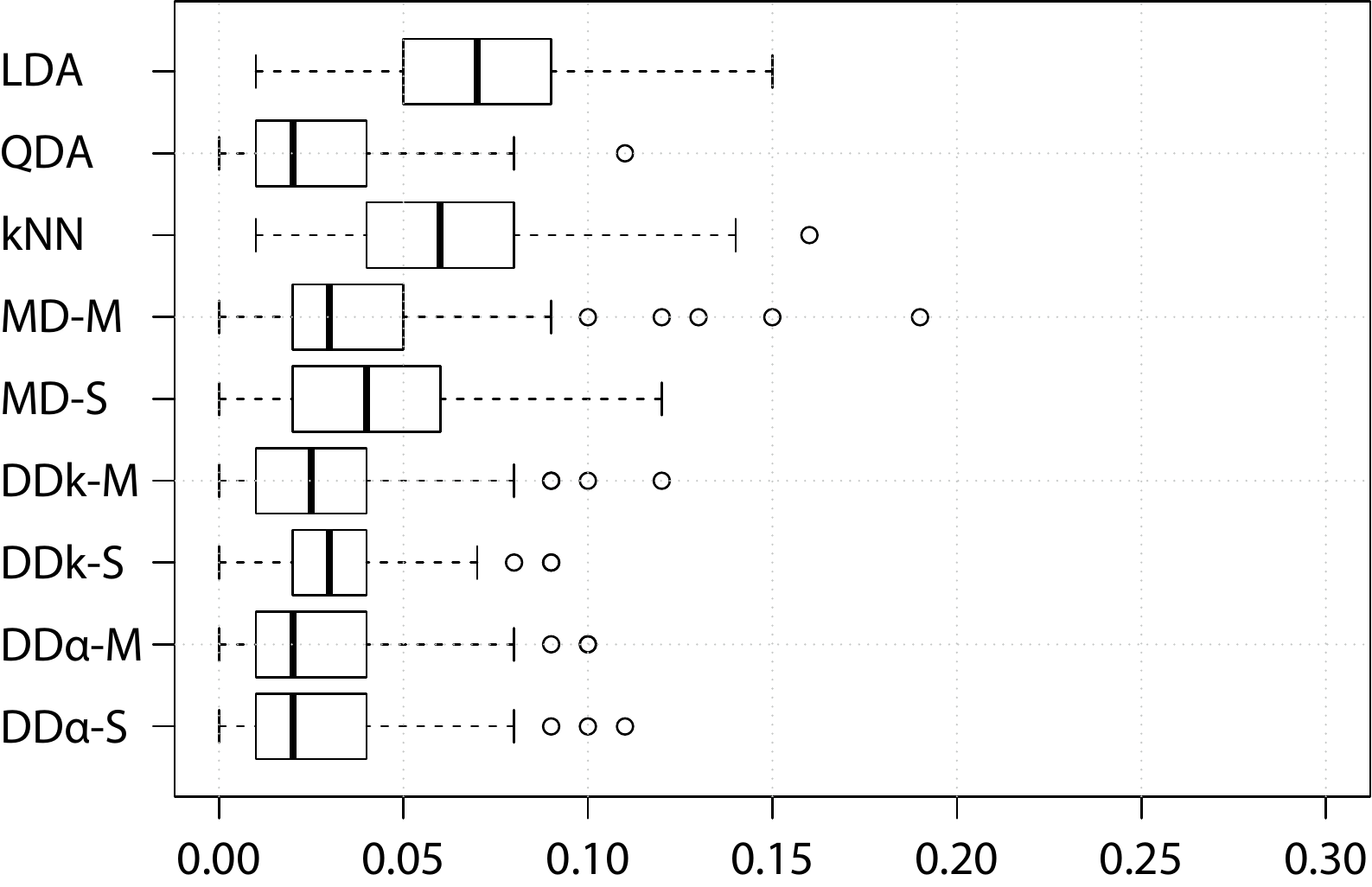}
	\caption{Boxplots of error rates for Model 1 (left) and Model 2 (right), using Vapnik-Chervonenkis bound (top), cross-validation (middle) and componentwise method (bottom).}
    \label{fig:simboxplot}
\end{figure*}

We take 100 observations (50 from each class) for training and 100 (50 from each class) for evaluating the performance of the classifiers. Training and classification are repeated 100 times to get stable results. Figure~\ref{fig:simboxplot} (left) presents boxplots (over 100 takes) of error rates of twelve classifiers applied after transforming the data to properly constructed finite-dimensional spaces.
The top panel refers to a location-slope space, where $L$ and $S$ are selected by Vapnik-Chervonenkis restricted cross-validation ({\em VCcrossLS}), the middle panel to a location-slope space whose dimensions are determined by  mere, unrestricted cross-validation ({\em crossLS}), the bottom panel to the
finite-dimensional argument subspace constructed by the componentwise method {\em crossDHB} of \citep{DelaigleHB12}.
The classifiers are: linear (LDA) and quadratic (QDA) discriminant analysis, $k$-nearest neighbors classifier ($kNN$), maximum depth classifier with Mahalanobis (MD-M), spatial (MD-S) and projection (MD-P) depth, $DD$-plot classifier with $kNN$ rule based on $L_\infty$ distance ($DDk$-M, $DDk$-S, $DDk$-P), and $DD\alpha$-classifier ($DD\alpha$-M, $DD\alpha$-S, $DD\alpha$-P), both with the three depths, correspondingly.
The last approach (\emph{crossDHB}) has not been combined with the projection depth for two reasons. First, performing the necessary cross-validations with the projection depth becomes computationally infeasible; for computational times
see Table~\ref{tab:timesim} in Appendix~B. Second, the quality of approximation of the projection depth differs between the tries, and this instability is possibly misleading when choosing the optimal argument subspace, thus yielding rather high error rates; compare, e.g., the classification errors for  \emph{growth data}, Table~\ref{tab:realerrors} in Section~\ref{ssec:benchmarks}.

As expected, all $DD$-plot-based classifiers, applied in a properly chosen location-slope space, show highly satisfactory performance, which is in line with the best result of \cite{CuevasFF07}. The location-slope spaces that have been selected for the different classifiers, \emph{viz.} the corresponding $(L,S)$-pairs, do not much differ; see Table~\ref{tab:model1dimsVCb} in Appendix~B.
The picture remains the same when the location-slope space is chosen using unrestricted cross-validation \emph{crossLS}, which yields no substantial improvement.

On the other hand, classifiers operating in an optimal argument subspace (\emph{crossDHB})
are outperformed by those employed in the location-slope space (\emph{crossLS, VCcrossLS}), although their error rates are still low; see Figure~\ref{fig:simboxplot} (left).
A plausible explanation could be that differences between the classes at each single argument point are not significant enough, and a bunch of them has to be taken for reasonable discrimination. But the sequential character of the procedure (as discussed in the Introduction and in Appendix~A)
prevents from choosing higher dimensions. Most often the dimensions two or three are chosen; see Table~\ref{tab:model1dimsCMP} in Appendix~B. Our procedure appears to be better, as by integrating more information is extracted from the functions, so that they become distinguishable.

Next we consider an example, where averaging comes out to be rather a disadvantage.
{\em Model 2} of \cite{CuevasFF07} looks as follows:
\[X_0=\{\mathbf{x}^0(t)|\mathbf{x}^0(t)=30(1-t)t^2+0.5|\sin(20\pi t)|+u(t)\}\]
with $u(t)$ and $M=51$ as before. $X_1$ is an 8-knot spline approximation of $X_0$. See Figure~\ref{fig:models.CFF07} (right) for illustration.

The corresponding boxplots of the error rates are depicted in Figure~\ref{fig:simboxplot} (right). The results for individual classifiers are different. When the location-slope space is chosen using Vapnik-Chervonenkis bound (\emph{VCcrossLS}), LDA, $kNN$ and all maximum depth classifiers perform poorly, while the $DDk$-classifiers with Mahalanobis and spatial depths perform better. The $DD\alpha$-classifiers perform comparably. The last two lack efficiency when employed with projection depth; as seen from Table~\ref{tab:model2dimsVCb} in Appendix~B, the efficient location-slope spaces are of substantially higher dimension (8 and more). Thus, the larger error rates with projection depth are explained by the insufficient number of random directions used in approximating this depth.
Also, with projection depth, different to Mahalanobis and spatial depth, less efficient $(L,S)$-pairs are preferred; see Tables
\ref{tab:model2dimsVCb} and \ref{tab:model2dimsCV} in Appendix~B.

Choosing the location-slope space by unrestricted cross-validation (\emph{crossLS}) does not change a lot.
The error rates obtained with this location-slope space are larger than those obtained with the synthesized space (\emph{crossDHB}),
although with QDA and $DD$-plot-based classifiers they stay reasonable. In Model 2, taking several extreme points would be enough for distinguishing the classes, and the finite-dimensional spaces have most often dimension four, and sometimes three. (Note, that all $DD$-plot-based classifiers regard these dimensions as sufficient, and, together with QDA, deliver best results.) On the other hand, the classifiers operating in some location-slope space select efficient dimension 8 and higher, which is also seen from Tables~\ref{tab:model2dimsVCb} and~\ref{tab:model2dimsCV} in Appendix~B.

\subsection{Comparisons on benchmark data}\label{ssec:benchmarks}
Now we come back to the two benchmark problems given in the Introduction.
The \emph{growth data} have been already analyzed by several authors.
\cite{LopezPR06} achieve a best classification error of {\bf 16.13}~\% when classifying with the $L_1$ distance from the trimmed mean and trimming is based on the generalized band depth with trimming parameter $\alpha=0.2$. \cite{CuestaANR10} use an average distance weighting with the random Tukey depth and get classification error {\bf 13.68}~\%.
\cite{CuevasFF07} obtain mean error rate of {\bf 9.04}~\% when using the double random projection depth and {\bf 4.04}~\% with $kNN$, dividing the sample into 70 training and 23 testing observations over 70 tries.
{\cite{CuestaAFBOF14} achieve {\bf 3.5}~\% when employing the LDA-classifier on a $DD$-plot that is based on the $h$-mode depth applied to both location and slope.}
\cite{SgueraGL13} get error rates of {\bf 3.45}~\% when classifying using kernelized functional spatial depth and choosing kernel parameter by means of cross-validation.

\begin{table}[t]
\centering
\caption{Error rates (in \%) when selecting a proper location-slope dimension.}
\label{tab:realerrors}
\begin{tabular}{|l|l|l|l|l|l|}
\hline   & \multicolumn{3}{|c|}{growth data} & \multicolumn{2}{|c|}{medflies data} \\ \hline
Data set     & {VCcrossLS} & {crossLS} & {crossDHB} & {VCcrossLS} & {crossLS} \\ \hline\hline
LDA          & 4.3   & 4.3   & 3.23  & 39.33 & 41.01 \\ \hline
QDA          & 4.3   & 5.38  & 7.53  & 38.58 & 42.7  \\ \hline
kNN          & 3.23  & 3.23  & 4.3   & 41.39 & 44.76 \\ \hline
MD-M         & 5.38  & 5.38  & 5.38  & 38.95 & 40.64 \\ \hline
MD-S         & 7.53  & 7.53  & 7.53  & 38.2  & 38.01 \\ \hline
MD-P         & 8.6   & 7.53  & 9.68  & 44.01 & 43.82 \\ \hline
$DDk$-M      & 5.38  & 5.38  & 5.38  & 43.26 & 41.39 \\ \hline
$DDk$-S      & 4.3   & 4.3   & 7.53  & 39.33 & 41.95 \\ \hline
$DDk$-P      & 5.38  & 3.23  & 8.6   & 42.32 & 41.57 \\ \hline
$DD\alpha$-M & 5.38  & 5.38  & 3.23  & 37.83 & 38.58 \\ \hline
$DD\alpha$-S & 5.38  & 5.38  & 6.45  & 38.95 & 40.26 \\ \hline
$DD\alpha$-P & 6.45  & 6.45  & 6.45  & 35.02 & 35.96 \\ \hline
\end{tabular}
\end{table}

Table~\ref{tab:realerrors} (columns \emph{growth data}) provides errors, estimated by leave-one-out cross-validation, of different classification techniques. In this, either a proper location-slope space is based on Vapnik-Chervonenkis bound (column \emph{VCcrossLS}), on
unrestricted cross-validation (\emph{crossLS}), or an optimal argument subset is chosen by the componentwise technique of \cite{DelaigleHB12} (\emph{crossDHB}).
Note that with \emph{VCcrossLS} classification by $kNN$ is best. It achieves error rate {\bf 3.23} \%, which means here that only three observations are misclassified. It coincides with a best cross-validation result of
\cite{BailloC08}. Both $DD$-plot-based classifiers perform well with all three depths, while the maximum depth classifiers MD-S and MD-P perform worse.

The Vapnik-Chervonenkis restricted cross-validation \emph{VCcrossLS} seems to perform not much worse than the unrestricted cross-validation \emph{crossLS}, and it mostly outperforms
the componentwise approach {\em crossDHB}. The latter is particularly bad when the projection depth is involved.
Tables~\ref{tab:growthdimsVCb},~\ref{tab:growthdimsCV} and~\ref{tab:growthdimsCMP} in Appendix~B exhibit how often various $(L,S)$-pairs and dimensions of optimal argument subspace are chosen.

In general, all three space-constructing techniques allow for very low error rates, producing as little as three misclassifications if the classifier is properly chosen. The $DD$-classifiers on $LS$-spaces yield at most six misclassifications.

Compared to \cite{LopezPR06} and \cite{CuestaANR10}, the better performance of our classifiers appears to be due to the inclusion of average slopes. Observe that the acceleration period starting with approximately nine years discriminates particularly well between girls and boys; see Figure~\ref{fig:growth2}. Note that also the componentwise method prefers (location) values from this interval.

A much more involved real task is the classification of  the \emph{medflies data}. In \cite{MuellerS05} these data are analyzed by generalized functional linear models. The authors employ logit regression and semi-parametric quasi-likelihood regression. They get errors of {\bf 41.76} \% and {\bf 41.2} \%, respectively, also estimated by leave-one-out cross-validation.

We apply all classifiers to the same data in properly constructed location-slope spaces. With our procedure we are able to improve the differential between long- and short-lived flies.
Especially, with the $DD\alpha$-classifier based on projection depth an error of {\bf 35.02} \% is obtained (see Table~\ref{tab:realerrors}, columns captioned \emph{medflies data} for the errors). The role of the derivatives in building the location-slope space is emphasized in Tables~\ref{tab:medfliesdimsVCb} and~\ref{tab:medfliesdimsCV} in Appendix~B, which show the frequencies at which the various $(L,S)$-pairs are selected. {\em crossLS}  is outperformed in most of the cases.
We were not able to compare the componentwise approach {\em crossDHB} as the computational load is too high.
On an average, $DD\alpha$-classifiers perform very satisfactory.
LDA and QDA with Vapnik-Chervonenkis bound, and maximum-depth classifiers with Mahalanobis and spatial depth (MD-M, MD-S), also deliver reasonable errors.

Note that, in configuring the location-slope space with \emph{crossLS}, lower errors could be obtained by using finer (e.g. leave-one-out) cross-validations.  To make componentwise classification feasible and the comparison fair, we have used only 10-fold cross-validation in all procedures besides $kNN$. ($k$ in $kNN$ and $DD$-plot-based $kNN$ has been determined by leave-one-out cross-validation.) For exact implementation details the reader is referred to Appendix~A.

\subsection{Computational loads}\label{ssec:time}
Most methods of functional classification tend to be time consuming because of their needs for preprocessing, smoothing and parameter-tuning, but the literature on such methods usually does not discuss computation times.
Nevertheless this is an important practical issue.
Our procedure comes out to be particularly efficient due to three main reasons.
Firstly, an eventual cross-validation is restricted to very few iterations.
Secondly, the depth space, where the final classification is done, has low dimension, which equals the number of classes.
Thirdly, the linear interpolation requires no preprocessing or smoothing of the data.

To illustrate this we briefly present the durations of both training and classification phases for the two real data sets and the twelve classification techniques in Table~\ref{tab:timereal}. (Classification time of a single object is reported in parentheses below.) As the times depend on implementation and platform used, we also report (in square brackets) the numbers of cross-validations done, as this measure is independent of the eventual classification technique.
The training times have been obtained as the average over all runs of leave-one-out cross-validation (thus using 92, respectively 533, observations for \emph{ growth} and \emph{medflies data}). This comes very close to the time needed to train with the entire data set, as the difference of one observation is negligible. The classification times in Table 2 have been obtained in the same way, i.e. averaging the classification of each single observation over all runs of the leave-one-out cross-validation. The same holds for the number of cross-validating iterations. For the componentwise classifiers ({\it crossDHB}) the averages have been replaced by the medians for the following reason. The sequential character of the procedure causes an exponential increase of time with each iteration (in some range; see Appendix~A for implementation details). Therefore, occasionally the total computation time can be outlying. (In our study, once the time exceeded four hours, {\em viz.} when classifying \emph{growth data} with the $DD$-plot-based $kNN$-classifier and projection depth, which required 20450 iterations to cross-validate.) On the other hand, when employing faster classifiers (which usually require stronger assumption on the data) the training phase can take less than two minutes. (This has been pointed by \cite{DelaigleHB12} as well.)

\begin{table}[!h]
\centering
\caption{Average (median for componentwise classification=crossDHB) training and classification (in parentheses) times (in seconds), and numbers of cross-validations performed (in square brackets), estimated by leave-one-out cross-validation.}
\label{tab:timereal}
{\footnotesize
\begin{tabular}{|l|l|l|l|l|l|}
\hline & \multicolumn{3}{|c|}{growth data} & \multicolumn{2}{|c|}{medflies data} \\ \hline
Data set     & {VCcrossLS} & {crossLS} & {crossDHB} & {VCcrossLS} & {crossLS} \\ \hline\hline
LDA          & 2.73     & 12.42    & 150.78    & 9.87     & 29.36     \\
             & (0.002)  & (0.0019) & (0.002)   & (0.0021) & (0.0023)  \\
             & [5.39]   & [150]    & [3110]    & [5]      & [150]     \\ \hline
QDA          & 2.73     & 12.23    & 48.45     & 9.77     & 29.12     \\
             & (0.0017) & (0.0017) & (0.0019)  & (0.0016) & (0.0027)  \\
             & [5.39]   & [150]    & [1020]    & [5]      & [150]     \\ \hline
kNN          & 2.91     & 26.38    & 213.36    & 23.87    & 814.46    \\
             & (0.001)  & (0.001)  & (0.0011)  & (0.0024) & (0.0028)  \\
             & [5.39]   & [150]    & [2576]    & [5]      & [150]     \\ \hline
MD-M         & 2.49     & 2.77     & 65.82     & 9.46     & 9.86      \\
             & (0.0009) & (0.0009) & (0.0004)  & (0.0007) & (0.0007)  \\
             & [5.39]   & [150]    & [6920]    & [5]      & [150]     \\ \hline
MD-S         & 2.63     & 5.47     & 168.14    & 10.04    & 35.89     \\
             & (0.0016) & (0.0017) & (0.0017)  & (0.0017) & (0.0018)  \\
             & [5.39]   & [150]    & [6940]    & [5]      & [150]     \\ \hline
MD-P         & 4.51     & 78.32    & 1795.86   & 19.86    & 439.03    \\
             & (0.0398) & (0.0481) & (0.0392)  & (0.2247) & (0.2233)  \\
             & [5.39]   & [150]    & [4699]    & [5]      & [150]     \\ \hline
$DDk$-M      & 3.06     & 17.31    & 299.3     & 20.53    & 335.48    \\
             & (0.0012) & (0.0011) & (0.0012)  & (0.002)  & (0.002)   \\
             & [5.39]   & [150]    & [2856]    & [5]      & [150]     \\ \hline
$DDk$-S      & 3.61     & 36.31    & 631.97    & 24.13    & 551.36    \\
             & (0.0022) & (0.0021) & (0.0019)  & (0.003)  & (0.0033)  \\
             & [5.39]   & [150]    & [3135]    & [5]      & [150]     \\ \hline
$DDk$-P      & 6.65     & 143.42   & 3103.
57   & 41.56    & 1145.16   \\
             & (0.0308) & (0.0308) & (0.03)    & (0.1995) & (0.1987)  \\
             & [5.39]   & [150]    & [4115]    & [5]      & [150]     \\ \hline
$DD\alpha$-M & 3.42     & 24.7     & 182.03    & 14.98    & 174.56    \\
             & (0.0009) & (0.0009) & (0.0011)  & (0.001)  & (0.001)   \\
             & [5.39]   & [150]    & [1020]    & [5]      & [150]     \\ \hline
$DD\alpha$-S & 4        & 42.49    & 860.14    & 18.71    & 392.61    \\
             & (0.0018) & (0.0017) & (0.0017)  & (0.0018) & (0.0019)  \\
             & [5.39]   & [150]    & [3135]    & [5]      & [150]     \\ \hline
$DD\alpha$-P & 7.02     & 154.24   & 2598.38   & 36.04    & 983.61    \\
             & (0.0306) & (0.0309) & (0.0298)  & (0.2)    & (0.1995)  \\
             & [5.39]   & [150]    & [3135]    & [5]      & [150]     \\ \hline
\end{tabular}
}
\end{table}

With \emph{growth data} training times are substantially higher when choosing an $(L,S)$-pair by unrestricted cross-validation than when the Vapnik-Chervonenkis bound is employed. Though, for the fast maximum depth classifier (with Mahalanobis or spatial depth) computation times appear quite comparable. Application of {\it crossDHB} causes an enormous increase in time (as well as in the number of cross-validations needed).
 For \emph{medflies data}, as expected, \emph{VCcrossLS} is faster than \emph{crossLS}. We were not able to perform the leave-one-out cross-validation estimation for the componentwise method for this data set, because of its excessive computational load.

See also Table~\ref{tab:timesim} for the same experiment regarding simulated data. Here, the projection-depth-based classifiers have not been implemented at all, as they need too much computation time.

\section{Conclusions}\label{sec:conclusions}
An efficient nonparametric procedure has been introduced for binary classification of functional data. The procedure consists of a two-step transformation of the original data  plus a classifier operating on the unit square. The functional data are first mapped into a finite-dimensional location-slope space and then transformed by a multivariate depth function into the $DD$-plot, which is a subset of the unit square. Three alternative depth functions are employed for this, as well as two rules for the final classification on $[0,1]^2$.

Our procedure either outperforms or matches existing approaches on simulated as well as on real benchmark data sets. The results of the $DD$-plot-based $kNN$ and the $DD\alpha$-procedure are generally good, although, (cf.\ Model 2) they are slightly outperformed by the componentwise classification method of \cite{DelaigleHB12}.

As the raw data are linearly interpolated, no spurious information is added.
The core of our procedure is the new data-dependent construction of the location-slope space. We bound its dimension $L+S$ by a Vapnik-Chervonenkis bound.
The subsequent depth transformation into the unit {square} makes the procedure rather robust since the final classification is done on a low-dimensional compact set.

Our use of statistical data depth functions demonstrates the variety of their application and opens new prospects when considering the proposed location-slope space. To reflect the dynamic structure of functional data, the construction of this space, in a natural way, takes levels together with derivatives into account. As it has been shown, efficient information extraction is done via piece-wise averaging of the functional data in its raw form, while the changing of the functions with their argument is reflected by their average slopes.

The finite-dimensional space has to be constructed in a way that respects the important intervals and includes most information. Here, equally spaced intervals  are used that cover the entire domain but have possibly different numbers for location and slope. This gives sufficient freedom in configuring the location-slope space.
Note, that in view of the simulations as well as the benchmark results, choosing a particular depth function is of limited relevance only.
While, depending on given data, different intervals are differently relevant,
location and slope may differ in information content as well.
The set of reasonable location-slope spaces is enormously reduced by application of the
Vapnik-Chervonenkis bound, and the selection is done by fast cross-validation over a very small set.
The obtained finite-dimensional space can be augmented
by coordinates reflecting additional information on the data, that may be available.
While our procedure is designed for the analysis of functions that are piecewise smooth, it can be also used for functions that are just piecewise continuous (by simply fixing the parameter $S$ at 0).
Obviously, higher order derivatives can be involved, too. But obtaining those requires smooth extrapolation, which affords additional computations and produces possibly spurious information.


The new approach is presented here for $q=2$ classes, but it is not limited to this case. If $q>2$, $kNN$ is applied in the $q$-dimensional depth space without changes, and the $DD\alpha$-classifier is extended by either constructing $q$ one-against-all or ${q}\choose{2}$ pair-wise classifiers in the depth space, and finally performing some aggregation in the classification phase; \citep[see also][]{LangeMM12a}.

Our space selection technique (incomplete cross-validation, restricted by a Vapnik-Chervonenkis bound) is
compared, both in terms of error rates and computational time, with a full cross-validation
as well as with the componentwise space synthesis method of \cite{DelaigleHB12}. We do this for all variants of the classifiers.


In future research comparisons with other existing functional classification techniques as well as the use of other finite-dimensional classifiers on the $DD$-plot are needed. Refined data-dependent procedures, which size the relevant intervals and leave out irrelevant ones, may be developed to configure the location-slope space. However such refinements will possibly conflict with the efficiency and generality of the present approach.

\section*{Acknowledgement}

We thank Dominik Liebl for his critical comments on an earlier version of the manuscript, as well as Ondrej Vencalek and Aurore Delaigle for their helpful remarks. The reading and suggestions of two referees are also gratefully acknowledged.





\appendix
\section{Implementation details}\label{sec:app1}
In calculating the depths, $\mu_Y$ and $\Sigma_Y$ for the \emph{Mahalanobis depth} have been determined by the usual moment estimates and similarly, $\Sigma_Y$ for the {\em spatial depth}. The {\em projection depth} has been approximated by drawing 1\,000 directions from the uniform distribution on the unit sphere. Clearly, the number of directions needed for satisfactory approximation depends on the dimension of the space. Observe that for higher-dimensional problems 1\,000 directions are not enough, which becomes apparent from the analysis of Model~2 in Section~\ref{ssec:simresults}. There the location-slope spaces chosen have dimension eight and higher; see also Tables~\ref{tab:model2dimsVCb} and~\ref{tab:model2dimsCV} in Appendix~B. On the other hand, calculating the projection depth even in one dimension costs something. Computing 1\,000 directions to approximate the projection depth takes substantially more time than computing the exact Mahalanobis  or spatial depths (see Tables~\ref{tab:timereal} and~\ref{tab:timesim} in Appendix~B).

{\em LDA} and {\em QDA} are used with classical moment estimates, and priors are estimated by the class portions in the training set.
The {\em $kNN$-classifier} is applied to location-slope data in its affine invariant form, based on the covariance matrix of the pooled classes. For time reasons, its parameter $k$ is determined by leave-one-out cross-validation over a reduced range, {\em viz.} $k\in \{1, \dots, \max\{\min\{10(m+n)^{1/d}+1,m+n-1\},2\}\}$.
The $\alpha$-procedure separating the $DD$-plot uses polynomial space extensions with maximum degree three; the latter is selected by cross-validation.
To keep the training speed of the depth-based $kNN$-classifier comparable with that of the $DD\alpha$-classifier, we also determine $k$ by leave-one-out cross-validation on a reduced range of $k\in \{1, \dots, \max\{\min\{10\sqrt{m+n}+1,(m+n)/2\},2\}\}$.

Due to \emph{linear interpolation}, the levels are integrated as piecewise-linear functions, and the derivatives as piecewise constant ones.
If the dimension of the location-slope space is too large (in particular for inverting the covariance matrix, as it can be the case in Model~2), PCA is used to reduce the dimension. Then $\epsilon_{max}$ is estimated and all further computations are performed in the subspace of principal components having positive loadings.

To \emph{construct the location-slope space}, firstly all
pairs $(L,S)$ satisfying $2\le L+S\le M/2$ are considered.
($M/2$ amounts to 26 for the synthetic and to 16 for the real data sets.)
For each $(L,S)$ the data are transformed to $\mathbb{R}^{L+S}$, and the Vapnik-Chervonenkis bound $\epsilon_{max}$ is calculated. Then those five pairs are selected that have smallest $\epsilon_{max}$. Here, tied values of $\epsilon_{max}$ are taken into account as well, with the consequence that on an average slightly more than five pairs are selected;
see the {\em growth data} in Table~\ref{tab:timereal} and both synthetic models in Table~\ref{tab:timesim} of Appendix~B. Finally, among these the best $(L,S)$-pair is chosen by means of cross-validation. Note that the goal of this cross-validation is not to actually choose a best location-slope dimension but rather to get rid of obviously misleading $(L,S)$-pairs, which may yield relatively small values of $\epsilon_{max}$. This is seen from Figures~\ref{fig:VapnikC1} and~\ref{fig:VapnikC2}.
When determining an optimal $(L,S)$-pair by \emph{crossLS}, the same set of $(L,S)$-pairs is considered as with \emph{VCcrossLS}.

In implementing the \emph{componentwise method} of finite-dimensional space synthesis (\emph{crossDHB}) we have
followed \cite{DelaigleHB12} with slight modifications.
The original approach of \cite{DelaigleHB12} is combined with the sequential approach of \cite{FerratyHV10}.
Initially, a grid of equally ($\Delta t$) distanced discretization points is built. Then a sequence of finite-dimensional spaces is synthesized  by adding points of the grid step by step.
We start with all pairs of discretization points that have at least distance $2\Delta t$. (Note that
\cite{DelaigleHB12} start with single points instead of pairs.)
The best of them is chosen by cross-validation. Then step by step features are added.
In each step, that point that has best discrimination power (again, in the sense of cross-validation) when added to the already constructed set is chosen as a new feature. The resulting set of points is used to construct a neighborhood of combinations to be further considered.
As a neighborhood we use twenty $2\Delta t$-distanced points in the second step, and ten in the third; from the fourth step on the
sequential approach is applied only.

All our \emph{cross-validations} are ten-fold, except the leave-one-out cross-validations in determining $k$ with both $kNN$-classifiers.
Of course, partitioning the sample into ten parts only may depreciate our approach against a more
comprehensive leave-one-out cross-validation.
We have chosen it to keep computation times of the {\em crossDHB} approach \citep{DelaigleHB12} in practical
limits and also to make the comparison of approaches equitable throughout our study.

The calculations have been implemented in an \emph{R-environment}, based on the R-package ``ddalpha'' \citep{MozharovskyiML13}, with speed critical parts written in C++. The R-code implementing our methodology as well as that performing the experiments can be obtained upon request from the authors.
In all experiments, one kernel of the processor Core\,i7-2600 (3.4\,GHz) having enough physical memory has been used.
Thus, regarding the methodology of \cite{DelaigleHB12} our implementation differs from their original one and, due to its module-based structure, may result in larger computation times.
For this reason we provide the number of cross-validations performed; see Tables~\ref{tab:timereal} and~\ref{tab:timesim} of Appendix~B.
The comparison appears to be fair, as we always use ten-fold cross-validation together with an identical set of classification rules in the finite-dimensional spaces.

\section{Additional tables}\label{sec:app2}
\begin{table}[h!]
\centering
\caption{Frequency (in \%) of selected location-slope dimensions using the Vapnik-Chervonenkis bound; Model~1.}
\label{tab:model1dimsVCb}
{\small
\begin{tabular}{|c|c|c|c|c|c|c|c|c|c|c|c|c|}
\hline  &  &  &  & \multicolumn{3}{|c|}{Max.depth} & \multicolumn{3}{|c|}{$DD$-$kNN$}  & \multicolumn{3}{|c|}{$DD\alpha$} \\ \cline{5-13}
$(L,S)$ & LDA   & QDA   & $kNN$ & Mah.  & Spt.  & Prj.  & Mah.  & Spt.  & Prj.  & Mah.  & Spt.  & Prj.             \\ \hline\hline
 (2,1)  & 65    & 64    & 55    & 58    & 63    & 54    & 65    & 63    & 47    & 49    & 51    & 50               \\ \hline 
 (3,1)  & 15    & 23    & 29    & 34    & 23    & 29    & 16    & 24    & 34    & 29    & 26    & 31               \\ \hline 
 (2,0)  & 14    &  7    &  7    &  3    &  9    & 11    &  7    &  3    &  5    & 11    & 12    & 10               \\ \hline 
 (2,2)  &  1    &  2    &  5    &  0    &  1    &  3    &  4    &  2    &  6    &  3    &  1    &  6               \\ \hline 
 (3,0)  &  2    &  1    &  4    &  1    &  1    &  1    &  2    &  1    &  3    &  4    &  5    &  3               \\ \hline 
 Others &  3    &  3    &  0    &  4    &  3    &  2    &  6    &  7    &  5    &  4    &  5    &  0               \\ \hline 
\end{tabular}
}
\end{table}

\begin{table}[h!]
\centering
\caption{Frequency (in \%) of selected location-slope dimensions using the Vapnik-Chervonenkis bound; Model~2.}
\label{tab:model2dimsVCb}
{\footnotesize
\begin{tabular}{|c|c|c|c|c|c|c|c|c|c|c|c|c|}
\hline  &  &  &  & \multicolumn{3}{|c|}{Max.depth} & \multicolumn{3}{|c|}{$DD$-$kNN$}  & \multicolumn{3}{|c|}{$DD\alpha$} \\ \cline{5-13}
$(L,S)$ & LDA   & QDA   & $kNN$ & Mah.  & Spt.  & Prj.  & Mah.  & Spt.  & Prj.  & Mah.  & Spt.  & Prj.             \\ \hline\hline
 (5,4)  & 41    & 51    & 27    & 39    & 31    &  0    & 40    & 37    &  0    & 37    & 37    &  1               \\ \hline 
 (0,8)  & 19    & 23    & 24    & 31    & 40    & 20    & 13    & 12    & 30    & 16    & 10    & 23               \\ \hline 
 (1,8)  &  4    &  4    & 13    &  9    & 10    & 23    &  8    &  6    & 18    &  3    &  4    & 21               \\ \hline 
 (2,8)  &  8    &  8    &  3    &  1    &  0    &  7    & 15    & 14    &  4    & 12    &  8    &  4               \\ \hline 
 (4,8)  &  1    &  4    &  5    &  1    &  1    &  6    &  6    &  9    &  4    &  8    &  6    &  6               \\ \hline 
 (3,8)  &  7    &  4    &  3    &  0    &  1    &  8    &  0    &  3    &  6    &  7    &  5    &  4               \\ \hline 
 (6,4)  &  2    &  2    &  4    &  3    &  3    &  0    &  7    &  6    &  0    &  8    & 12    &  0               \\ \hline 
 (6,8)  &  3    &  2    &  1    &  1    &  0    &  4    &  1    &  1    &  6    &  0    &  3    &  4               \\ \hline 
 (5,8)  &  2    &  0    &  1    &  0    &  1    &  3    &  2    &  3    &  7    &  0    &  4    &  2               \\ \hline 
 (10,8) &  2    &  1    &  3    &  2    &  1    &  5    &  0    &  0    &  5    &  0    &  2    &  2               \\ \hline 
 (7,8)  &  2    &  0    &  0    &  0    &  0    &  0    &  2    &  0    &  6    &  3    &  0    &  5               \\ \hline 
 (12,8) &  1    &  0    &  1    &  1    &  0    &  5    &  1    &  1    &  3    &  0    &  0    &  3               \\ \hline 
 (18,8) &  1    &  0    &  0    &  0    &  0    &  2    &  0    &  0    &  0    &  0    &  0    &  5               \\ \hline 
 Others &  7    &  1    & 15    & 12    & 12    & 17    &  5    &  8    & 11    &  6    &  9    & 20               \\ \hline 
\end{tabular}
}
\end{table}

\begin{table}[h!]
\centering
\caption{Frequency (in \%) of location-slope dimensions chosen using the Vapnik-Chervonenkis bound; growth data.}
\label{tab:growthdimsVCb}
{\footnotesize
\begin{tabular}{|c|c|c|c|c|c|c|c|c|c|c|c|c|}
\hline  &  &  &  & \multicolumn{3}{|c|}{Max.depth} & \multicolumn{3}{|c|}{$DD$-$kNN$}  & \multicolumn{3}{|c|}{$DD\alpha$} \\ \cline{5-13}
$(L,S)$ & LDA   & QDA   & $kNN$ & Mah.  & Spt.  & Prj.            & Mah.  & Spt.  & Prj.              & Mah.  & Spt.  & Prj.             \\ \hline\hline
 (0,2)  & 87.1  & 94.62 & 64.52 & 46.24 & 45.16 & 23.66           & 27.96 &  3.23 & 41.94             & 72.04 & 66.67 & 36.56            \\ \hline %
 (1,2)  & 12.9  &  5.38 & 35.48 &  6.45 & 19.35 &  0              & 17.20 & 89.25 & 32.26             & 24.73 & 26.88 & 39.78            \\ \hline %
 (0,3)  &  0    &  0    &  0    & 24.73 &  4.30 & 68.82           & 27.96 &  2.15 &  3.23             &  0    &  0    &  1.08            \\ \hline %
 (2,2)  &  0    &  0    &  0    & 21.51 & 30.11 &  1.08           &  1.08 &  3.23 & 11.83             &  1.08 &  0    & 17.20            \\ \hline %
 (1,3)  &  0    &  0    &  0    &  0    &  1.08 &  1.08           & 23.66 &  2.15 &  9.68             &  0    &  5.38 &  4.30            \\ \hline %
 (4,0)  &  0    &  0    &  0    &  1.08 &  0    &  0              &  2.15 &  0    &  1.08             &  2.15 &  1.08 &  1.08            \\ \hline %
 (0,4)  &  0    &  0    &  0    &  0    &  0    &  5.38           &  0    &  0    &  0                &  0    &  0    &  0               \\ \hline %
\end{tabular}
}
\end{table}

\begin{table}[h!]
\centering
\caption{Frequency (in \%) of location-slope dimensions chosen using the Vapnik-Chervonenkis bound; medflies data.}
\label{tab:medfliesdimsVCb}
{\footnotesize
\begin{tabular}{|c|c|c|c|c|c|c|c|c|c|c|c|c|}
\hline  &       &       &       & \multicolumn{3}{|c|}{Max.depth} & \multicolumn{3}{|c|}{$DD$-$kNN$}  & \multicolumn{3}{|c|}{$DD\alpha$} \\ \cline{5-13}
$(L,S)$ & LDA   & QDA   & $kNN$ & Mah.  & Spt.  & Prj.            & Mah.  & Spt.  & Prj.              & Mah.  & Spt.  & Prj.             \\ \hline\hline
 (1,1)  &  1.87 &   0   &  3.18 &  0.75 & 93.45 & 80.34           & 11.42 &  8.05 & 70.04             & 74.91 & 87.64 & 98.13            \\ \hline %
 (2,1)  & 84.27 &   0   & 63.48 &  0    &  0    &  0              & 64.42 & 79.96 &  0.19             & 24.53 & 12.17 &  0               \\ \hline %
 (0,2)  &  9.18 & 100   & 32.58 & 99.25 &  6.55 & 18.54           & 23.97 & 10.11 & 27.34             &  0.56 &  0.19 &  1.31            \\ \hline %
 (1,2)  &  1.12 &   0   &  0    &  0    &  0    &  1.12           &  0.19 &  0.19 &  2.43             &  0    &  0    &  0.56            \\ \hline %
 (3,1)  &  3.56 &   0   &  0.75 &  0    &  0    &  0              &  0    &  1.69 &  0                &  0    &  0    &  0               \\ \hline %
\end{tabular}
}
\end{table}

\begin{table}[h!]
\centering
\caption{Frequency (in \%) of selected location-slope dimensions using cross-validation; Model~1.}
\label{tab:model1dimsCV}
{\footnotesize
\begin{tabular}{|c|c|c|c|c|c|c|c|c|c|c|c|c|}
\hline  &  &  &  & \multicolumn{3}{|c|}{Max.depth} & \multicolumn{3}{|c|}{$DD$-$kNN$}  & \multicolumn{3}{|c|}{$DD\alpha$} \\ \cline{5-13}
$(L,S)$ & LDA   & QDA   & $kNN$ & Mah.  & Spt.  & Prj.  & Mah.  & Spt.  & Prj.  & Mah.  & Spt.  & Prj.             \\ \hline\hline
 (2,1)  & 47    & 49    & 34    & 43    & 41    & 31    & 45    & 42    & 35    & 50    & 42    & 31               \\ \hline 
 (3,1)  & 14    & 21    & 17    & 11    & 22    & 15    & 13    & 14    & 14    & 17    & 12    & 15               \\ \hline 
 (4,1)  &  8    &  1    &  7    &  6    &  4    &  6    &  1    &  7    &  3    &  3    &  7    &  8               \\ \hline 
 (2,0)  &  6    &  2    &  3    &  5    &  5    &  1    &  5    &  1    &  5    &  3    &  4    &  4               \\ \hline 
 (2,2)  &  0    &  1    &  9    &  1    &  2    &  4    &  6    &  5    &  1    &  5    &  4    &  2               \\ \hline 
 (5,1)  &  1    &  0    &  2    &  6    &  4    &  3    &  2    &  3    &  2    &  1    &  1    &  3               \\ \hline 
 (2,3)  &  2    &  1    &  6    &  2    &  4    &  0    &  1    &  4    &  0    &  2    &  3    &  0               \\ \hline 
 (3,2)  &  0    &  4    &  5    &  3    &  1    &  3    &  3    &  2    &  2    &  0    &  2    &  0               \\ \hline 
 (3,0)  &  1    &  0    &  2    &  0    &  0    &  1    &  3    &  0    &  2    &  1    &  1    &  1               \\ \hline 
 Others & 21    & 21    & 15    & 23    & 17    & 36    & 21    & 22    & 36    & 18    & 24    & 36               \\ \hline 
\end{tabular}
}
\end{table}

\begin{table}[h!]
\centering
\caption{Frequency (in \%) of selected location-slope dimensions using cross-validation; Model~2.}
\label{tab:model2dimsCV}
{\footnotesize
\begin{tabular}{|c|c|c|c|c|c|c|c|c|c|c|c|c|}
\hline  &  &  &  & \multicolumn{3}{|c|}{Max.depth} & \multicolumn{3}{|c|}{$DD$-$kNN$}  & \multicolumn{3}{|c|}{$DD\alpha$} \\ \cline{5-13}
$(L,S)$ & LDA   & QDA   & $kNN$ & Mah.  & Spt.  & Prj.  & Mah.  & Spt.  & Prj.  & Mah.  & Spt.  & Prj.             \\ \hline\hline
 (5,4)  & 38    & 56    & 27    & 42    & 35    &  0    & 41    & 47    &  0    & 44    & 36    &  0               \\ \hline 
 (0,8)  & 17    &  6    & 14    & 21    & 33    &  6    &  5    &  2    & 11    &  3    &  5    & 14               \\ \hline 
 (6,4)  &  6    &  7    & 11    &  3    &  5    &  0    & 11    & 13    &  0    & 10    & 17    &  0               \\ \hline 
 (10,0) &  0    & 11    &  0    &  0    &  0    &  3    & 17    & 15    &  0    & 20    & 16    &  0               \\ \hline 
 (2,8)  &  4    &  8    &  4    &  2    &  0    &  8    &  6    & 14    &  1    & 11    & 11    &  8               \\ \hline 
 (1,8)  &  7    &  2    &  6    &  9    &  9    & 10    &  2    &  1    & 11    &  1    &  1    & 10               \\ \hline 
 (3,8)  &  7    &  1    &  9    &  0    &  0    &  8    &  3    &  2    &  8    &  1    &  0    &  7               \\ \hline 
 (4,8)  &  1    &  1    &  2    &  0    &  0    &  9    &  5    &  1    &  8    &  1    &  8    &  6               \\ \hline 
 (5,8)  &  1    &  0    &  2    &  1    &  0    &  9    &  0    &  0    &  4    &  1    &  0    &  9               \\ \hline 
 (6,8)  &  1    &  0    &  3    &  1    &  2    &  8    &  0    &  0    &  3    &  0    &  1    &  6               \\ \hline 
 (7,8)  &  0    &  0    &  1    &  0    &  1    &  5    &  0    &  0    &  8    &  0    &  0    &  6               \\ \hline 
 (9,0)  &  0    &  6    &  0    &  0    &  0    &  0    &  5    &  2    &  0    &  2    &  2    &  0               \\ \hline 
 (5,7)  &  2    &  0    &  2    &  7    &  5    &  0    &  0    &  0    &  0    &  0    &  0    &  0               \\ \hline 
 (16,8) &  0    &  0    &  1    &  0    &  0    &  5    &  0    &  0    &  5    &  0    &  0    &  4               \\ \hline 
 (12,8) &  1    &  0    &  1    &  0    &  0    &  1    &  0    &  0    &  7    &  0    &  0    &  4               \\ \hline 
 (8,8)  &  3    &  0    &  1    &  0    &  0    &  2    &  0    &  0    &  6    &  0    &  0    &  1               \\ \hline 
 (17,8) &  0    &  0    &  0    &  0    &  0    &  2    &  0    &  0    &  3    &  0    &  0    &  8               \\ \hline 
 (13,8) &  0    &  0    &  1    &  0    &  0    &  3    &  0    &  0    &  2    &  0    &  0    &  6               \\ \hline 
 (10,8) &  2    &  0    &  0    &  0    &  0    &  0    &  0    &  0    &  6    &  0    &  0    &  3               \\ \hline 
 (18,8) &  0    &  0    &  0    &  0    &  0    &  5    &  0    &  0    &  3    &  0    &  0    &  1               \\ \hline 
 Others & 10    &  2    & 15    & 14    & 10    & 16    &  5    &  3    & 14    &  6    &  3    &  7               \\ \hline 
\end{tabular}
}
\end{table}

\begin{table}[h!]
\centering
\caption{Frequency (in \%) of location-slope dimensions chosen using cross-validation; growth data.}
\label{tab:growthdimsCV}
{\scriptsize
\begin{tabular}{|c|c|c|c|c|c|c|c|c|c|c|c|c|}
\hline  &  &  &  & \multicolumn{3}{|c|}{Max.depth} & \multicolumn{3}{|c|}{$DD$-$kNN$}  & \multicolumn{3}{|c|}{$DD\alpha$} \\ \cline{5-13}
$(L,S)$ & LDA   & QDA   & $kNN$ & Mah.  & Spt.  & Prj.            & Mah.  & Spt.  & Prj.              & Mah.  & Spt.  & Prj.             \\ \hline\hline
 (0,2)  & 87.1  & 93.55 & 64.52 & 44.09 & 44.09 &  0              & 25.81 &  3.23 & 19.35             & 68.82 & 60.22 &  3.23            \\ \hline 
 (1,2)  & 12.9  &  5.38 & 35.48 &  6.45 & 19.35 &  0              & 17.20 & 70.97 & 12.90             & 24.73 & 25.81 & 17.20            \\ \hline 
 (0,3)  &  0    &  0    &  0    & 21.51 &  4.30 & 29.03           & 26.88 &  2.15 &  1.08             &  0    &  0    &  1.08            \\ \hline 
 (2,2)  &  0    &  0    &  0    & 11.83 & 29.03 &  1.08           &  1.08 &  2.15 &  8.60             &  1.08 &  0    &  7.53            \\ \hline 
 (4,0)  &  0    &  0    &  0    & 13.98 &  2.15 &  0              &  5.38 &  1.08 &  4.30             &  5.38 & 10.75 &  5.38            \\ \hline 
 (1,3)  &  0    &  0    &  0    &  0    &  1.08 &  0              & 21.51 &  2.15 &  5.38             &  0    &  2.15 &  0               \\ \hline 
 (4,2)  &  0    &  0    &  0    &  0    &  0    &  0              &  0    &  7.53 &  1.08             &  0    &  0    &  9.68            \\ \hline 
 (2,3)  &  0    &  0    &  0    &  0    &  0    &  1.08           &  2.15 &  5.38 &  1.08             &  0    &  0    &  1.08            \\ \hline 
 (4,5)  &  0    &  0    &  0    &  0    &  0    &  0              &  0    &  0    &  2.15             &  0    &  0    &  7.53            \\ \hline 
 (4,6)  &  0    &  0    &  0    &  0    &  0    &  0              &  0    &  0    &  0                &  0    &  0    &  6.45            \\ \hline 
 (7,1)  &  0    &  0    &  0    &  0    &  0    &  5.38           &  0    &  0    &  0                &  0    &  0    &  0               \\ \hline 
 Others &  0    &  1.07 &  0    &  2.14 &  0    & 63.43           &  0    &  5.36 & 44.08             &  0    &  1.07 & 40.84            \\ \hline 
\end{tabular}
}
\end{table}

\begin{table}[h!]
\centering
\caption{Frequency (in \%) of location-slope dimensions chosen using cross-validation; medflies data.}
\label{tab:medfliesdimsCV}
{\scriptsize
\begin{tabular}{|c|c|c|c|c|c|c|c|c|c|c|c|c|}
\hline  &       &        &       & \multicolumn{3}{|c|}{Max.depth} & \multicolumn{3}{|c|}{$DD$-$kNN$}  & \multicolumn{3}{|c|}{$DD\alpha$} \\ \cline{5-13}
$(L,S)$ & LDA   & QDA    & $kNN$ & Mah.  & Spt.  & Prj.            & Mah.  & Spt.  & Prj.              & Mah.  & Spt.  & Prj.             \\ \hline\hline
 (1,1)  &  0    &   0    &  0.75 &  0.56 & 89.89 & 77.72           &  0.37 &  0    & 64.98             & 71.54 & 79.59 & 98.31            \\ \hline 
 (0,2)  &  7.68 &   0    & 14.23 & 91.39 &  4.12 & 19.66           &  1.87 &  0    & 24.34             &  0.37 &  0.19 &  1.69            \\ \hline 
 (2,1)  & 57.87 &   0    & 47.00 &  0    &  0    &  0              &  0    &  1.87 &  0                &  0    &  0    &  0               \\ \hline 
 (0,6)  &  0    &   0    &  4.68 &  0    &  0    &  0              & 32.02 & 14.23 &  5.24             &  0.37 &  7.49 &  0               \\ \hline 
 (2,1)  &  0    &   0    &  0    &  0    &  0    &  0              &  6.55 &  0    &  0                & 22.28 & 10.86 &  0               \\ \hline 
 (7,5)  & 30.90 &   0    &  0    &  0    &  0    &  0              &  0    &  0    &  0                &  0    &  0    &  0               \\ \hline 
 (2,6)  &  0    &   0    & 21.16 &  0    &  0    &  0              &  0.19 &  0.75 &  0                &  0    &  0    &  0               \\ \hline 
 (14,2) &  0    &  18.35 &  0    &  0    &  0    &  0              &  0.75 &  2.06 &  0                &  0    &  0.19 &  0               \\ \hline 
 (2,12) &  0    &   9.36 &  0    &  0    &  0    &  0              &  3.75 &  3.75 &  0                &  0    &  0    &  0               \\ \hline 
 (13,3) &  0    &  12.55 &  0    &  0    &  0    &  0              &  0.94 &  1.69 &  0                &  0.56 &  0    &  0               \\ \hline 
 (1,12) &  0    &   0    &  0    &  0    &  0    &  0              & 12.17 &  3.37 &  0.19             &  0    &  0    &  0               \\ \hline 
 (3,9)  &  0    &   0.37 &  0    &  0    &  0    &  0              &  7.49 &  6.74 &  0                &  0    &  0    &  0               \\ \hline 
 (7,6)  &  0    &   3.93 &  0    &  0    &  0    &  0              &  3.18 &  7.30 &  0                &  0    &  0    &  0               \\ \hline 
 (11,2) &  0    &   8.24 &  0    &  0    &  0    &  0              &  3.56 &  1.50 &  0                &  0.56 &  0    &  0               \\ \hline 
 (0,5)  &  0    &   0    &  0    &  7.49 &  5.99 &  0              &  0    &  0.19 &  0                &  0    &  0    &  0               \\ \hline 
 (3,12) &  0    &   0    &  0    &  0    &  0    &  0              &  4.49 &  8.61 &  0                &  0    &  0    &  0               \\ \hline 
 (4,3)  &  0    &   0    &  0.19 &  0    &  0    &  0              &  0.56 & 11.80 &  0                &  0    &  0    &  0               \\ \hline 
 (7,3)  &  0    &  10.49 &  0    &  0    &  0    &  0              &  0.94 &  0    &  0                &  0    &  0    &  0               \\ \hline 
 (4,6)  &  0    &  10.86 &  0    &  0    &  0    &  0              &  0    &  0.19 &  0                &  0    &  0    &  0               \\ \hline 
 (11,5) &  0    &   6.55 &  0    &  0    &  0    &  0              &  0.19 &  1.50 &  0                &  0    &  0    &  0               \\ \hline 
 (4,10) &  0    &   0    &  0    &  0    &  0    &  0              &  2.81 &  5.06 &  0                &  0    &  0    &  0               \\ \hline 
 (15,0) &  0    &   6.74 &  0    &  0    &  0    &  0              &  0.37 &  0.19 &  0                &  0    &  0    &  0               \\ \hline 
 (1,3)  &  0    &   0    &  0.19 &  0    &  0    &  0              &  0    &  6.74 &  0.19             &  0    &  0    &  0               \\ \hline 
 Others &  3.55 &  12.56 & 11.80 &  0.56 &  0    &  2.62           & 17.80 & 22.46 &  5.06             &  5.44 &  0.56 &  0               \\ \hline 
\end{tabular}
}
\end{table}

\begin{table}[h!]
\centering
\caption{Frequency (in \%) of selected dimensions using componentwise method; Model~1.}
\label{tab:model1dimsCMP}
{\footnotesize
\begin{tabular}{|c|c|c|c|c|c|c|c|c|c|}
\hline  &  &  &  & \multicolumn{2}{|c|}{Max.depth} & \multicolumn{2}{|c|}{$DD$-$kNN$}  & \multicolumn{2}{|c|}{$DD\alpha$} \\ \cline{5-10}
 dim    & LDA    & QDA    & $kNN$  & Mah.   & Spt.   & Mah.   & Spt.   & Mah.   & Spt.              \\ \hline\hline
  2     &  50    &  55    &  45    &  45    &  50    &  46    &  47    &  50    &  64               \\ \hline %
  3     &  41    &  38    &  46    &  46    &  43    &  48    &  40    &  42    &  27               \\ \hline %
  4     &   9    &   7    &   9    &   9    &   7    &   6    &  13    &   8    &   9               \\ \hline %
\end{tabular}
}
\end{table}

\begin{table}[h!]
\centering
\caption{Frequency (in \%) of selected dimensions using componentwise method; Model~2.}
\label{tab:model2dimsCMP}
{\footnotesize
\begin{tabular}{|c|c|c|c|c|c|c|c|c|c|}
\hline  &  &  &  & \multicolumn{2}{|c|}{Max.depth} & \multicolumn{2}{|c|}{$DD$-$kNN$}  & \multicolumn{2}{|c|}{$DD\alpha$} \\ \cline{5-10}
 dim    & LDA    & QDA    & $kNN$  & Mah.   & Spt.   & Mah.   & Spt.   & Mah.   & Spt.              \\ \hline\hline
  3     &  11    &  14    &  14    &   8    &  14    &  20    &  13    &  17    &  18               \\ \hline %
  4     &  89    &  86    &  51    &  54    &  60    &  79    &  85    &  83    &  81               \\ \hline %
  5     &   0    &   0    &  26    &  35    &  23    &   1    &   2    &   0    &   1               \\ \hline %
  6     &   0    &   0    &   9    &   3    &   2    &   0    &   0    &   0    &   0               \\ \hline %
  7     &   0    &   0    &   0    &   0    &   1    &   0    &   0    &   0    &   0               \\ \hline %
\end{tabular}
}
\end{table}

\begin{table}[h!]
\centering
\caption{Frequency (in \%) of selected dimensions using componentwise method; growth data.}
\label{tab:growthdimsCMP}
{\footnotesize
\begin{tabular}{|c|c|c|c|c|c|c|c|c|c|c|c|c|}
\hline  &  &  &  & \multicolumn{3}{|c|}{Max.depth} & \multicolumn{3}{|c|}{$DD$-$kNN$}  & \multicolumn{3}{|c|}{$DD\alpha$} \\ \cline{5-13}
 dim    & LDA    & QDA    & $kNN$  & Mah.   & Spt.   & Prj.   & Mah.   & Spt.   & Prj.   & Mah.   & Spt.   & Prj.              \\ \hline\hline
  2     & 100.00 &  92.47 &  45.16 &  12.90 &   3.23 &  41.94 &  88.17 &  50.54 &  39.78 & 100.00 &  96.77 &  60.22            \\ \hline %
  3     &   0    &   6.45 &  30.11 &  62.37 &  60.22 &  52.69 &   7.53 &  44.09 &  53.76 &   0    &   2.15 &  38.71            \\ \hline %
  4     &   0    &   1.08 &  24.73 &  24.73 &  36.56 &   5.38 &   4.30 &   5.38 &   6.45 &   0    &   1.08 &   1.08            \\ \hline %
\end{tabular}
}
\end{table}

\begin{table}[h!]
\centering
\caption{Average (median for componentwise classification=crossDHB) training and classification (in parentheses) times (in seconds), and numbers of cross-validations performed (in square brackets), over 100 tries.}
\label{tab:timesim}
{\footnotesize
\begin{tabular}{|l|l|l|l|l|l|l|}
\hline & \multicolumn{3}{|c|}{Model 1} & \multicolumn{3}{|c|}{Model 2} \\ \hline
Data set     & VCcrossLS & crossLS & crossDHB  & VCcrossLS & crossLS & crossDHB \\ \hline\hline
LDA          & 7.61     & 40.59    & 547.5     & 6.83     & 30.3     & 392.16   \\
             & (0.0011) & (0.0011) & (0.001)   & (0.0012) & (0.0012) & (0.001)  \\
             & [5.55]   & [375]    & [10759.5] & [8.58]   & [375]    & [7648.5] \\ \hline
QDA          & 7.59     & 39.44    & 503.5     & 6.74     & 29.72    & 411.23   \\
             & (0.0011) & (0.0011) & (0.001)   & (0.0011) & (0.0011) & (0.001)  \\
             & [5.53]   & [375]    & [9628]    & [8.32]   & [375]    & [8009]   \\ \hline
kNN          & 7.86     & 133.77   & 1263.76   & 7.63     & 63.16    & 489.18   \\
             & (0.0012) & (0.0012) & (0.0011)  & (0.0013) & (0.0014) & (0.0011) \\
             & [5.27]   & [375]    & [12996]   & [8.8]    & [375]    & [4886]   \\ \hline
MD-M         & 7.38     & 7.53     & 101.05    & 6.22     & 7.11     & 58.45    \\
             & (0.0011) & (0.0011) & (0.001)   & (0.0011) & (0.0011) & (0.001)  \\
             & [5.46]   & [375]    & [10344.5] & [9.05]   & [375]    & [5884.5] \\ \hline
MD-S         & 7.48     & 16.98    & 259.75    & 6.38     & 14.36    & 169.54   \\
             & (0.0012) & (0.0012) & (0.0011)  & (0.0012) & (0.0012) & (0.0011) \\
             & [5.51]   & [375]    & [10113]   & [8.35]   & [375]    & [6518.5] \\ \hline
MD-P         & 9.58     & 245.52   &           & 10.76    & 195.61   &          \\
             & (0.0017) & (0.0018) & -         & (0.002)  & (0.0021) & -        \\
             & [5.56]   & [375]    &           & [8.96]   & [375]    &          \\ \hline
$DDk$-M      & 7.99     & 50.69    & 1421.2    & 7.26     & 48.95    & 601.2    \\
             & (0.0012) & (0.0013) & (0.0011)  & (0.0013) & (0.0013) & (0.001)  \\
             & [5.49]   & [375]    & [11909.5] & [9.1]    & [375]    & [5491]   \\ \hline
$DDk$-S      & 8.63     & 119.84   & 2493.52   & 8.71     & 102.75   & 1296.86  \\
             & (0.0013) & (0.0014) & (0.0012)  & (0.0014) & (0.0014) & (0.0012) \\
             & [5.44]   & [375]    & [10873]   & [9.68]   & [375]    & [5715.5] \\ \hline
$DDk$-P      & 12.16    & 453.28   &           & 14.42    & 383.15   &          \\
             & (0.0016) & (0.0016) & -         & (0.0017) & (0.0018) & -        \\
             & [5.62]   & [375]    &           & [8.06]   & [375]    &          \\ \hline
$DD\alpha$-M & 8.13     & 34.55    & 1866.99   & 7.93     & 76.58    & 995.68   \\
             & (0.0012) & (0.0012) & (0.001)   & (0.0012) & (0.0012) & (0.001)  \\
             & [5.57]   & [375]    & [10113]   & [9.43]   & [375]    & [5182]   \\ \hline
$DD\alpha$-S & 8.68     & 104.71   & 2840.91   & 9.06     & 128.62   & 1860.12  \\
             & (0.0012) & (0.0013) & (0.0011)  & (0.0013) & (0.0013) & (0.0012) \\
             & [5.51]   & [375]    & [9774]    & [8.99]   & [375]    & [6124]   \\ \hline
$DD\alpha$-P & 12.19    & 466.76   &           & 16.02    & 410.66   &          \\
             & (0.0015) & (0.0016) & -         & (0.0017) & (0.0017) & -        \\
             & [5.45]   & [375]    &           & [8.92]   & [375]    &          \\ \hline
\end{tabular}
}
\end{table}

\clearpage





\end{document}